\documentclass[%
 aip,jcp,floatfix,
 amsmath,amssymb,
 reprint,%
]{revtex4-1}
\usepackage[dvips]{graphicx}
\usepackage[dvipsnames]{xcolor}
\usepackage{soul}
\usepackage{epsfig}
\usepackage{float}
\usepackage{dcolumn}
\usepackage{bm}
\usepackage{amsmath}
\usepackage{amssymb}
\usepackage{hyperref}
\usepackage{natbib}
\usepackage[normalem]{ulem}
\usepackage[mathlines]{lineno}
\usepackage{placeins}
\usepackage[caption=false,font=normalsize,labelfont=bf]{subfig}
\definecolor{bostonuniversityred}{rgb}{0.8, 0.0, 0.0}

\begin{document}

\preprint{AIP/123-QED}

\title{Thermodynamics and kinetics of crystallisation in deeply supercooled Stillinger-Weber silicon} 

\author{Yagyik Goswami}
\affiliation{Theoretical Sciences Unit, Jawaharlal Nehru Centre for Advanced Scientific Research, Bengaluru, India.}
\author{Vishwas V. Vasisht}
\affiliation{Indian Institute of Technology Palakkad, Ahalia Integrated Campus, Kozhippara P.O. — Palakkad, India}
\author{Daan Frenkel}
\affiliation{Department of Chemistry, University of Cambridge, Cambridge, England}
\author{Pablo G. Debenedetti}
\affiliation{Department of Chemical and Biological Engineering, Princeton University, Princeton, NJ, USA}
\author{Srikanth Sastry}
\email{sastry@jncasr.ac.in}
\affiliation{Theoretical Sciences Unit, Jawaharlal Nehru Centre for Advanced Scientific Research, Bengaluru, India.}

\begin{abstract}

We study the kinetics of crystallisation in deeply supercooled liquid silicon employing computer simulations and the Stillinger-Weber three body potential. The free energy barriers to crystallisation are computed using umbrella sampling Monte Carlo simulations, and for selected low temperature and zero pressure state points, using unconstrained molecular dynamics simulations to reconstruct the free energy from a mean first passage time formulation. We focus on state points that have been 
described in earlier work [Sastry and Angell, Nature Mater., {\bf 2}, 739, 2003]
as straddling a first order liquid-liquid phase transition (LLPT) between two metastable liquid states.
It was argued subsequently [Ricci et al., Mol. Phys., {\bf 117}, 3254, 2019] that the apparent phase transition is in fact due the loss of metastability of the liquid state with respect to the globally stable crystalline state. The presence or absence of a barrier to crystallisation for these state points is therefore of importance to ascertain, with due attention to ambiguities that may arise from the choice of order parameters. 
We discuss our choice of order parameters and also our choice of methods to calculate the free energy at deep supercooling. 
We find a well-defined free energy barrier to crystallisation and demonstrate that both umbrella sampling and mean first passage time methods yield results that agree quantitatively.  Our results thus provide strong evidence against the possibility that the liquids at state points close to the reported LLPT exhibit slow, spontaneous crystallisation, but they do not address the existence of a LLPT (or lack thereof).
We also compute the free energy barriers to crystallisation at other state points over a broad range of temperatures and pressures, and discuss the effect of changes in the 
microscopic structure of the metastable liquid on the free energy barrier heights.  
\end{abstract} 

\maketitle

\section{Introduction}
The phase behaviour of liquid silicon is a subject of continuing interest due to the many anomalous properties it exhibits
reminiscent of water and of other tetrahedral liquids. Of particular interest is deeply supercooled silicon, i.e., the liquid cooled to temperatures significantly lower than the melting temperature. Here, as in water, anomalous behaviour such as a density maximum and the possibility of the existence of a first order phase transition between two metastable liquid states -- a high density liquid (HDL) and a low density liquid (LDL) -- has been the subject of numerous investigations that have approached the question from different directions\cite{vasisht2013liquid,tanaka2020liquid,sastry2003liquid,vasisht2011liquid,ricci2019computational,ashwin2004metal,jakse2007liquid,ganesh2009liquid,dharma2020liquid,aptekar1979phase,kim2005situ,beye2010liquid,zhang2014liquid,zhang2015polymorphism}.
The existence of a first order transition between ``amorphous" and liquid states was first proposed based on experimental observations\cite{spaepen_AIP_1979,bagley_AIP_1979,donovan1983heat,donovan1989homogeneous} 
and the possibility of a liquid-liquid transition was suggested on the basis of a simple two state model by Aptekar \cite{aptekar1979phase}. Notable experimental works since then, probing the phase behaviour of deeply supercooled silicon include the work of Kim {\it et al}\cite{kim2005situ}, where electrostatic levitation was used to prevent crystallisation induced by the container walls and temperatures as low as $T=1350K$ were probed. Subsequently, Beye {\it et al}\cite{beye2010liquid} used ultra-fast pump probe spectroscopy to discern changes in the electronic structure to identify a two-step change in the melt from semi-conductor to semi-metal to a high temperature metallic liquid.

Owing to the difficulties of conducting experiments on liquid silicon at these temperatures, as well as the difficulties in avoiding crystallisation \cite{shao1998analysis,hedler2004amorphous,kim2005situ,kim2008structural,watanabe2007does,beye2010liquid,okada2012persistence,okada2020phase},
computer simulations have played a significant role in efforts to study the liquid-liquid transition in silicon \cite{vasisht2013liquid,tanaka2020liquid,sastry2003liquid,jakse2007liquid,ganesh2009liquid,vasisht2011liquid,limmer2013putative,zhang2014liquid,zhang2015polymorphism,lu2015exploring,ricci2019computational,dharma2020liquid}. 
A number of simulation studies, including some of the most recent investigations in this area, employ {\it ab-initio} methods and identify liquid-liquid and liquid-solid transitions based on changes in the electronic structure reminiscent of those found in experiments; silicon is a semiconductor in the solid-state, a semi-metal in the low density liquid state and a metallic liquid in the high temperature, high density liquid state \cite{ashwin2004metal,jakse2007liquid,jakse2008dynamic,ganesh2009liquid,ganesh2011first,zhang2014liquid,zhao2015nature,remsing2018refined,dharma2020liquid}.

 Classical simulations using the Stillinger-Weber (SW) potential\cite{stillinger1985computer} have been performed extensively to probe relevant time scales whereby the metastable liquid phase can be studied in order to explore the possibility of a liquid-liquid transition \cite{vasisht2013liquid,sastry2010illuminating,sastry2003liquid,vasisht2011liquid,desgranges2011role,limmer2013putative,zhang2014liquid,lu2015exploring,ricci2019computational}.
At the relevant temperatures and pressures, the dynamics of the metastable liquid is sufficiently slow (relaxation times of tens of nanoseconds and longer) to make computer simulations challenging. On the other hand, crystal nucleation occurs on comparable time scales making experimental studies challenging. Employing simulations of SW silicon, Sastry and Angell~\cite{sastry2003liquid} observed a discontinuous change in enthalpy below the melting temperature, suggesting a first order phase transition between two states that were identified to be liquid-like based on structural and dynamical properties. Vasisht {\it et al}, 2011\cite{vasisht2011liquid}, identified a co-existence region and a transition line that ended at a critical point at negative pressures. These works estimated the transition temperature to be $\sim 1060K$ at $P=0~GPa$. Vasisht {\it et al}\cite{vasisht2011liquid} further illustrated the behaviour of important thermodynamic loci consistent with their observation of approach to a second critical point, similar to a number of models of water~\cite{poole1992phase,palmer2018advances,debenedetti2020second}. 

The question of the existence of two metastable liquid states for supercooled SW silicon has since been investigated through attempts to construct two dimensional free energy surfaces that may display distinct minima corresponding to the two liquid phases in addition to that corresponding to the stable crystal phase. Studies by Limmer and Chandler\cite{limmer2011putative,limmer2013putative} and by Ricci {\it et al}\cite{ricci2019computational}, evaluating the free energy surfaces, did not find any evidence of a metastable LDL. In fact, it was argued in these works that the metastable liquid was no longer stable with respect to crystallisation at the state points where earlier studies had found evidence of an LDL phase, and that crystallisation was spontaneous. In the context of water, a coarse-grained model of water based on reparametrising the SW model was employed to argue that increased crystallisation rates precluded the possibility of a transformation to the low density liquid phase~\cite{moore2011structural}, consistently with the above arguments. Nevertheless, from simulations of more explicit multi-site models of water such as the ST2, TIP4P and TIP5P models, and for the single-site spherically symmetric Jagla ramp potential, 
clear evidence of an LLPT ending at a critical point has been shown, notably in  \cite{abascal2010widom,kesselring2013finite,palmer2014metastable,ricci2017free,debenedetti2020second} among others. 

In the case of SW silicon, the claim that no free energy barrier separates the liquid sate from the crystal free energy minimum for state points in the vicinity of  $\sim 1060K$ at $P=0~GPa$ is puzzling, given the long simulation times over which the simulated systems have been observed in the liquid state~\cite{sastry2003liquid,vasisht2011liquid,vasisht2014nesting,angell2016potential}. 
A possible origin of such inconsistency is that the order parameters chosen to construct the free energy surfaces in ~\cite{limmer2011putative,limmer2013putative,ricci2019computational} lead to artefacts in the presence of low barriers to crystallisation, as briefly discussed in \cite{ricci2019computational}. In particular, the choice of a global order parameter ($Q_6$) as a measure of the degree of crystalline order may not permit a reversible control of crystallisation with the bias potentials used in umbrella sampling simulations. Related considerations with respect to the use of the global order parameter ($Q_6$) for evaluating free energy barriers have already been noted\cite{wolde1996simulation}. 

In the present work, we address one aspect of the issues surrounding the possibility of a liquid-liquid transition in SW silicon. As the discussion above makes clear, crystal nucleation rates play a central role, and among the possibilities that cast doubt on the possibility of the liquid-liquid transition, the most extreme case is that the liquid is simply not stable in the relevant state points, and crystal nucleation is spontaneous, or barrierless.
Thus, the first question that needs to be addressed is whether the liquid state is metastable, and hence finite free energy barriers to crystallisation exist, for the relevant state points. If the liquid state can be demonstrated to be metastable, one must address the separate question of whether two forms of the liquid exist, which we do not address in this work. 

In order to reliably compute free energy barriers to crystallisation, we need to also demonstrate that no artefacts arise as a result of the choice of order parameters in constrained simulations such as umbrella sampling. 
To this end, at deeply supercooled conditions, we compute the free energy profile for crystallisation using two independent methods, namely, (i) kinetic reconstruction of the free energy from unbiased molecular dynamics (MD) runs in the constant temperature, pressure and number of particles (NPT) ensemble, using the method described by Wedekind {\it et al}~\cite{wedekind2007new,wedekind2008kinetic,wedekind2009crossover} and (ii) Umbrella Sampling Monte Carlo simulations (USMC) in the NPT ensemble\cite{torrie1977nonphysical}, specifically, the prescription described by Saika-Voivod, Poole and Bowles,~\cite{saika2006test}. Both of these works have focused on cases of low free energy barriers and have discussed the specific considerations that become relevant to accurately measure them.

We find that finite free energy barriers and  well-defined critical nuclei, albeit small, exist for all the state points we investigate. We also demonstrate that the free energy profiles obtained using two independent methods agree well with each other for the state points considered. Thus, our results rule out the possibility that the liquid state is not stable for the range of state points across which a liquid-liquid transition has previously been claimed to arise. 

The rest of the paper is organised as follows: Section~\ref{sec:methods} describes the model potential used, the three-body Stillinger-Weber potential, the order parameters and the free energy calculation methods used. Section~\ref{sec:results} shows results obtained using the USMC simulations and from the kinetic reconstruction of the free energy from MD runs. A comparison of the free energy profiles is made.  Finally, a discussion of the results, ongoing work and outstanding issues follows in Section~\ref{sec:discussion}.


\section{Model and Methods} \label{sec:methods}

In this section we briefly describe the model potential and methods used in this study. A detailed description and discussion of the same is provided in Appendix~\ref{sec:Appendix_Model_and_Methods}.
We use the classical three-body Stillinger-Weber potential to model silicon~\cite{stillinger1985computer}. The model is designed to favour local tetrahedral ordering through the three-body interaction term and is the most widely used classical model of silicon.
In order to identify crystalline particles and crystalline clusters, we use the local analogue of the Steinhardt-Nelson bond orientational order parameters~\cite{steinhardt1983bond}.
The local bond ordering, typically denoted $q_l$, is calculated for each particle. Here, we use $q_3(i)$, noting that $q_6(i)$ can be used equivalently~\cite{vasisht2013phase,vasisht2014nesting}.
The $q_3(i)$ gives information about the ordering of the neighbours around the particle $i$. To determine bulk crystalline particles, we first identify particles with similarly ordered neighbourhoods by calculating $q_3(i).q_3(j)$.
Two particles are said to be ``bonded" if $Re(q_3(i).q_3(j))<-0.23$ and a bulk crystalline particle is one which has $q_3(i)>0.6$ and is bonded to at least 3 of its neighbours~\cite{van1992computer,ten1995numerical,wolde1996simulation,romano2011crystallisation,kesselring2013finite}. Two bulk crystalline atoms that are within the Stillinger-Weber cut-off distance, $3.78\AA$, of each other are said to belong to the same cluster. We consider both the largest cluster, denoted $n_{max}$, as well as the number of clusters of a given size $n$, denoted $N(n)$.

Free energy reconstructions are performed using two independent methods at low temperatures, along the $P=0GPa$ isobar, in order to obtain reliable estimates of the free energy barriers. The first method we employ is a kinetic reconstruction using the mean first passage time (MFPT) from unconstrained MD runs~\cite{wedekind2007new,wedekind2008kinetic,wedekind2009crossover}. In this method, the steady state probability of $n_{max}$, $P_{st}(n_{max})$, as well as the mean first passage time, $\tau_{MFPT}(n_{max})$ are computed from a collection of independent, crystallising trajectories and used to reconstruct the free energy using Eq.~\ref{eq:bdgvn} and Eq.~\ref{eq:Bn}. Further details are contained in Appendix~\ref{subsec:Appendix_MFPT}.
\begin{equation}
\beta \Delta G(x) = \beta \Delta G(x=1) + ln \left ( \frac{B(x)}{B(1)}\right ) - \int_{1}^{x} \frac{dx'}{B(x')}
\label{eq:bdgvn}
\end{equation}
\begin{equation}
B(x) = -\frac{1}{P_{st}(x)}\left [ \int_{x}^{b} P_{st}(x')dx' - \frac{\tau(b)-\tau(x)}{\tau(b)}\right].
\label{eq:Bn}
\end{equation}
Here, $x$ is the order parameter, which in this context is the size of the largest crystalline cluster, $n_{max}$. $b$ is the size of the largest crystalline cluster at which an absorbing boundary is imposed. $\Delta G(x)$ is the free energy of forming a crystalline nucleus of size $x$.
In order to compute the free energy using this method, $600$ independent NPT MD of $N=512$ particles simulations were started from disordered configurations with no crystalline particles and allowed to crystallise. The MD runs are performed on the LAMMPS software suite using the velocity Verlet algorithm with a timestep of $0.3830$ fs\cite{plimpton1995fast}. Thermostatting and barostatting are done with a Nos$\acute{e} $-Hoover thermostat/barostat with time constants of $100$ and $1000$ steps respectively.
\par
The other technique used to construct the free energies is umbrella sampling Monte Carlo~\cite{torrie1977nonphysical}. Simulations are performed in the NPT ensemble with constraints applied on the size of the largest crystalline cluster, $n_{max}$. Two bias potentials are used, a harmonic bias and a hard wall bias~\cite{saika2006test}. Parallel tempering swaps between simulations adjacent in temperature or bias potential are performed to speed up equilibration. For simulations where a hard wall bias is used, we begin simulations by applying a harmonic bias potential for $10^7$ MC steps before switching the bias potential. The auto-correlation functions of density ($\rho$), $Q_6$ and potential energy were monitored under the application of the hard wall bias and the relaxation time found to be similar and less than $10^5$ MC steps for all the windows and for each of the three quantities considered. Keeping in mind a relaxation time of $\tau=10^5$ MC steps, we use an equilibration length, under application of hard wall bias, of $50\tau$ and a production length of $250\tau$. We note here that the thermodynamic stability of the liquid is determined by whether there is a non-zero free energy cost to form small crystalline clusters which is maximum for some critical cluster size $n^{*}>0$. In using $n_{max}$ as the order parameter, we presume that $P(n_{max})~\approx~P(n)$, which is not necessarily true for small cluster sizes, particularly at low temperatures\cite{wolde1996simulation}. Further $P(n_{max})$ is expected to show a system size dependence while $P(n)$ is not. 
The statistics of the largest cluster, $n_{max}$ reveal that configurations containing a small cluster (i.e., where the largest cluster is small) are more frequently sampled than configurations where there are no crystalline clusters at all. This leads to the appearance of an artificial minimum in $\beta\Delta G(n_{max})$ at small values of $n_{max}$.
This issue has been discussed in the literature~\cite{saika2006test,chakrabarty2008chakrabarty,maibaum2008comment,wedekind2009crossover,lundrigan2009test} and a more extensive discussion is also included in Appendix~\ref{subsec:nmax-v-n}. Thus, in simulations where the hardwall bias is used, we gather statistics on the number of clusters of size $n$, $N(n)$. The quantity $P(n) = N(n)/N(0)$ can be related to the free energy as $\beta\Delta G(n) = -ln\left(P(n)\right)$ without the need to determine any additive constant since the way in which $P(n)$ is defined applies the constraint that $\beta\Delta G(n=0)=0$. To obtain statistics for the smallest cluster sizes, we perform simulations with a hardwall bias and use the full cluster size distribution to compute the free energy.

When reconstructing the free energy using either umbrella sampling or the kinetic reconstruction with $n_{max}$ as the order parameter we additionally specify that the free energy as a function of the largest cluster size, $\beta\Delta G(n_{max})$ be equal to $-ln\left(P(n)\right)$, for small cluster sizes.
By doing this, one obtains an estimate that can be meaningfully compared with $\beta\Delta G(n)$, the free energy from the full cluster size distribution. Similar techniques have been used in Ref.~\onlinecite{wedekind2009crossover} and in Ref.~\onlinecite{lundrigan2009test}.

We also perform free energy reconstructions with the global $Q_6$ as the order parameter 
(see Appendix~\ref{section:Appendix_Q6_MFPT_US} for the definition of $Q_6$), using both of the methods described above, and conclude that it is not a reliable order parameter to use to estimate the barrier to crystallisation. Details are contained in Appendix~\ref{section:Appendix_Q6_MFPT_US}.


\section{Results} \label{sec:results}
The results of free energy calculations performed at different state points are shown in this section, with a specific focus on temperatures across the previously reported LLPT at $P=0~GPa$~\cite{vasisht2011liquid}. At these temperatures, free energy calculations are performed using both the kinetic reconstruction from the MFPT and using umbrella sampling with a hard wall bias. The results for these state points are compared, showing a free energy barrier to the crystallisation transition at all the temperatures considered, demonstrating that crystallisation is not spontaneous.
The rest of the results are subsequently presented showing the free energy cost to crystallisation at other state points where the question of loss of metastability of the liquid does not arise. This includes free energy calculations performed at higher temperatures along the $P=0~GPa$ isobar. Calculations at low temperatures along the $P=0.75~GPa$ isobar are also performed. Along other isobars, the choice of state points is restricted to those understood to correspond to the high density liquid, based on the results in Ref.~\onlinecite{vasisht2011liquid}.
The free energy curves are also constructed along lines of constant coordination number, $C_{NN}$, and of constant isothermal compressibility, $\kappa_{T}$, in an attempt to understand the effect of density fluctuations and of the degree of tetrahedral ordering in the metastable liquid on the barrier to crystallisation. A further set of calculations is performed, crossing the line of maximum compressibility, known as the Widom line~\cite{xu2005relation,abascal2010widom}, beyond the LLCP reported in~\cite{vasisht2011liquid}, where the two purported metastable liquids cease to be indistinguishable.

Results for the different sets of state points are now presented in turn, after first illustrating the methodology for the treatment of free energy profiles at small cluster sizes, and the MFPT method.
We then discuss briefly the relationship between the free energy barrier and the critical nucleus size, and a comparison with the expectation based on classical nucleation theory (CNT).
Finally, for the low temperature $T = 1055K$ at zero pressure, we consider whether the choice of the initial ensemble of configurations (HDL-like or LDL-like) will make a difference to the estimation of free energy barriers, and answer it in the negative. 

\subsection{Comparing results at small cluster sizes}\label{subsec:compare}
\begin{figure}[htpb!]
\centering
\includegraphics[scale=0.5]{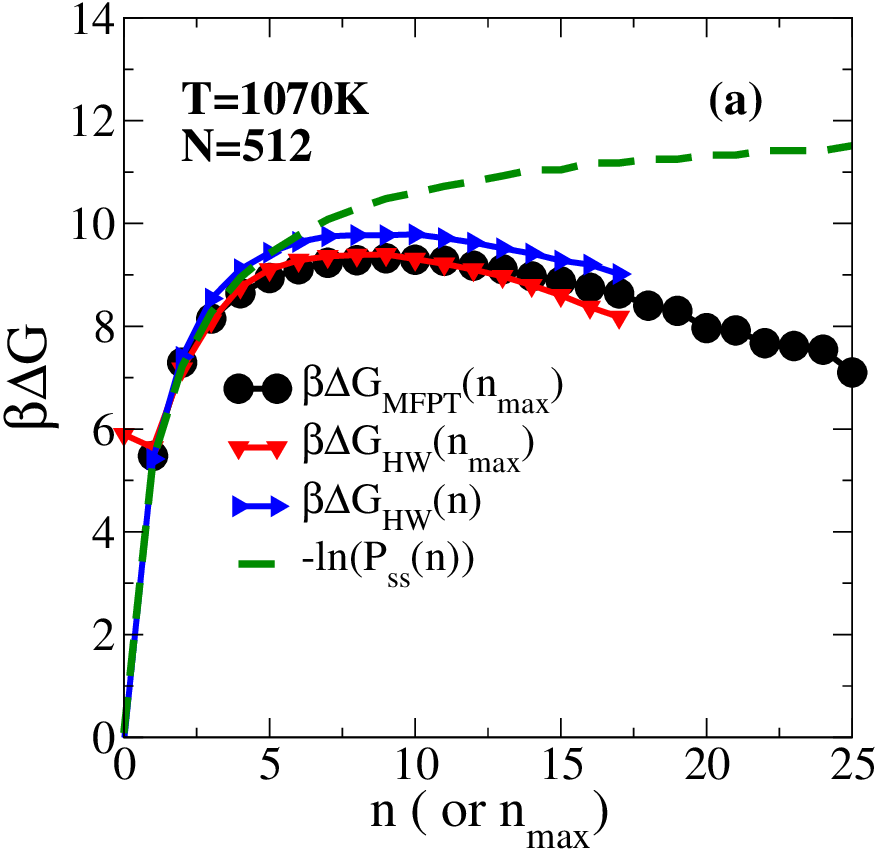}
\includegraphics[scale=0.5]{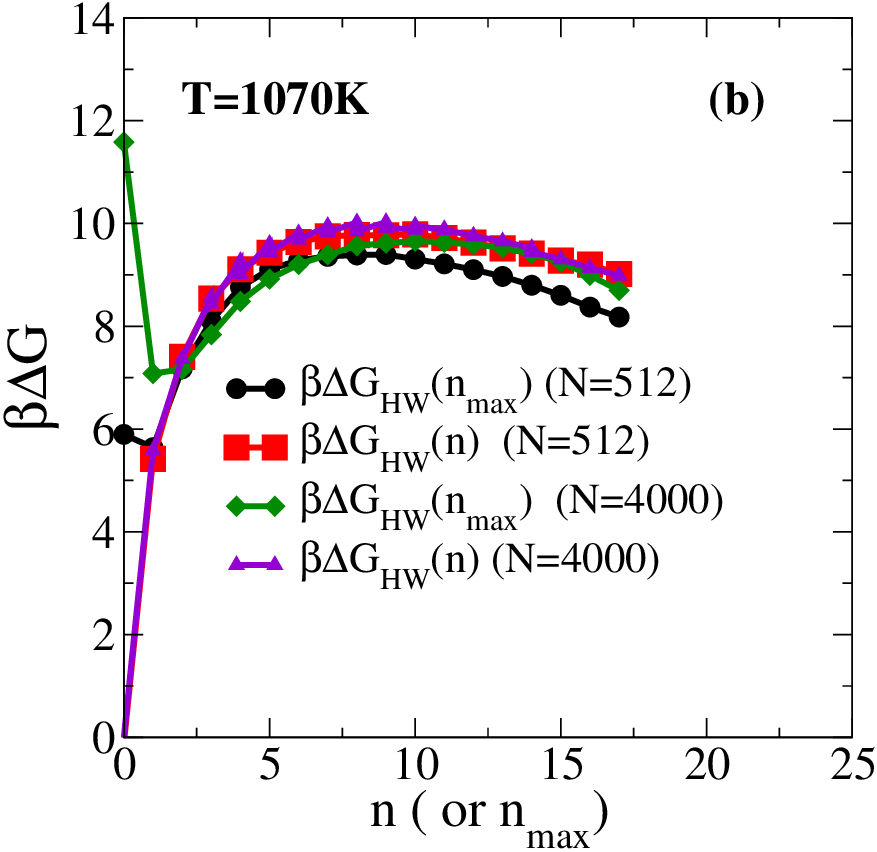}
\caption{A comparison of free energy reconstructions at $T=1070K$, $P=0~GPa$ using both $n$ and $n_{max}$ as order parameters at two system sizes, $N=4000$ and $N=512$. {\bf (a)} Comparison, at $N=512$, of results from the MFPT method using $n_{max}$ as the order parameter with results from the hard wall bias umbrella sampling using either $n$ or $n_{max}$ as the order parameter. {\bf (b)} Comparison of results using either $n$ or $n_{max}$ as the order parameter from the umbrella sampling simulations at two system sizes, $N=512$ and $N=4000$. For the purpose of comparison of $\beta\Delta G(n)$ with $\beta\Delta G(n_{max})$, the error in Eq.~\ref{eq:HW_mfpt_compare} is minimised. For $N=4000$, the error is minimised for $1~<~n \leq 3$.} 
\label{fig:fig-1}
\end{figure}

Following the procedure described in detail in Appendix~\ref{subsec:nmax-v-n}, we make a comparison between $\beta\Delta G(n)$ obtained from umbrella sampling runs with a hard wall bias to $\beta\Delta G(n_{max})$ obtained from both the kinetic reconstruction and umbrella sampling runs with a hard wall bias.
At $T=1070K$, $P=0~GPa$ (see Fig.~\ref{fig:fig-1}), using $n_{low}=n_{hi}=1$ gives nearly exact quantitative agreement between $\beta\Delta G(n)$ and $\beta\Delta G(n_{max})$ at $N=512$ regardless of the method used to generate the curves. As expected, free energy curves constructed from USMC simulations using the equilibrium $P(n)$ show no system-size dependence. In Fig.~\ref{fig:fig-1} {\bf (b)}, $\beta\Delta G_{HW}(n_{max})$ for small $n_{max}$ from umbrella sampling runs for $N=4000$ is obtained using $n_{low}=3,~n_{hi}=5$ (see Appendix~\ref{subsec:nmax-v-n},  Eq.~\ref{eq:HW_mfpt_compare}).
At lower temperatures, or even larger system sizes, as the appropriate value of $n_{low}$ becomes larger a comparison between $\beta\Delta G(n_{max})$ and $\beta \Delta G(n)$ can no longer be meaningfully made.
Having described how to compare the free energy results using the two methods, a comparison is made at $P=0~GPa$ at temperatures where the crystallisation transition is of particular interest. These results are shown in Section.~\ref{subsec:compare_results}.
\subsection{Kinetic reconstruction of free energy from MFPT}
\begin{figure}[htp!]
\centering
\includegraphics[scale=0.5]{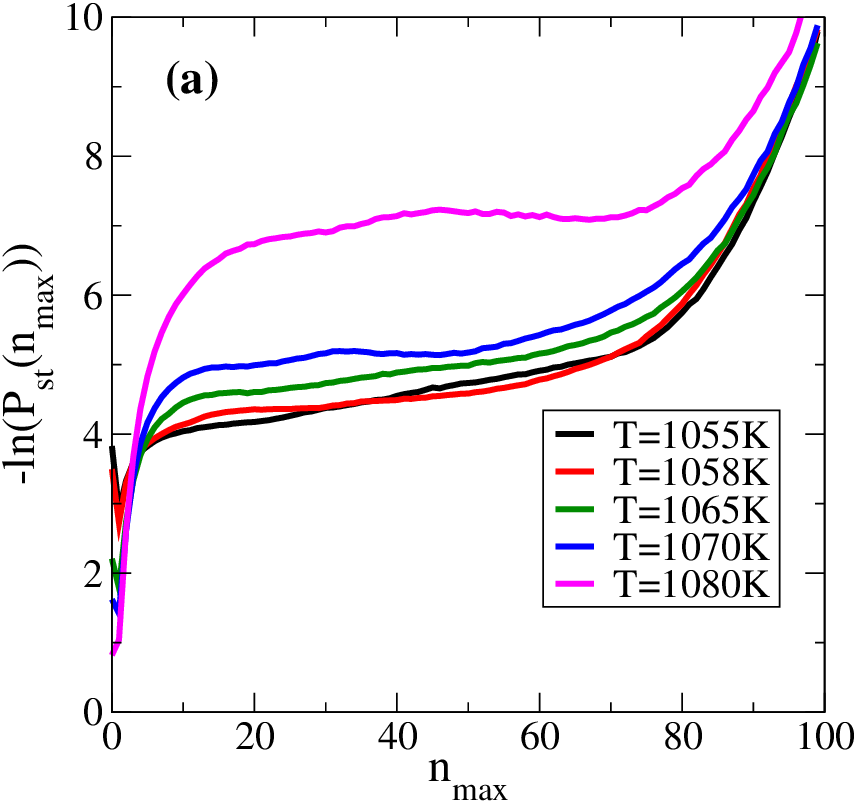}
\includegraphics[scale=0.5]{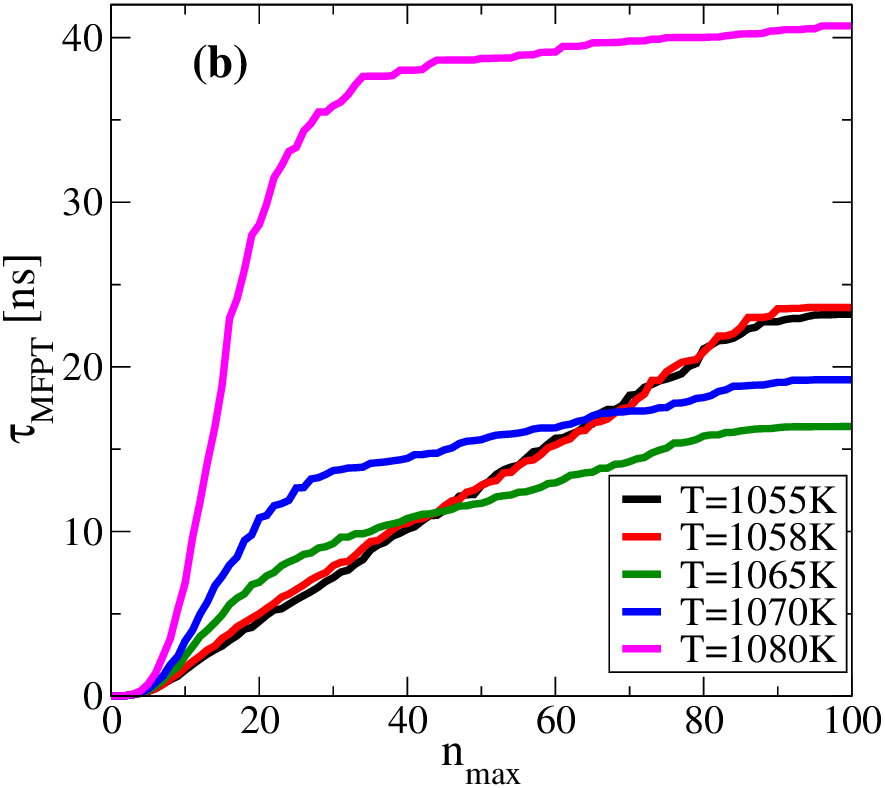}
\caption{{\bf (a)} $-ln(P_{st}(n_{max}))$ plotted against $n_{max}$ for different temperatures. $600$ NPT MD simulations at $P=0GPa$ were run with a system size of $N = 512$. {\bf (b) } $\tau_{MFPT}(n_{max})$ plotted against $n_{max}$ from the same set of MD runs.}
\label{fig:fig-2}
\end{figure}

The two main ingredients to reconstruct the free energy using this method are the MFPT, $\tau_{MFPT}(n_{max})$, and the steady state size distribution of the largest crystalline cluster, $P_{st}(n_{max})$. These can be used as shown in Eqs.~\ref{eq:bdgvn} and ~\ref{eq:Bn} to get the free energy with the largest cluster size as the order parameter, $\beta\Delta G(n_{max})$. The MFPT and steady state probability are shown in Fig.~\ref{fig:fig-2} for the temperatures studied here. These results are generated from NPT MD runs of $N=512$ particles. Results using this method are produced at state points where the pressure is $P=0~GPa$ and the temperature is varied in a range from high temperatures where the liquid can be unambiguously sampled in equilibrium before nucleating ($T=1070~K,~1080~K$) to lower temperatures where the loss of liquid metastability with respect to crystallisation becomes a consideration ($T~ < ~1070~K$). The order parameter is the size of the largest cluster, $n_{max}$ and the absorbing boundary condition is placed at $n_{max}=100$.
In Fig.~\ref{fig:fig-2}, we see that the MFPT, $\tau_{MFPT}(n_{max})$, shows a progressively decreasing sigmoidal character as we decrease the temperature from $T=1080K$ to $T=1055K$. This suggests that the difference between the nucleation timescale and the timescale of cluster growth decreases.

As discussed in Appendix~\ref{subsec:nmax-v-n}, the steady state probability $P_{st}(n_{max})$ shown in Fig.~\ref{fig:fig-2} {\bf (a)} peaks at small values of $n_{max}$ ( which shows up as a minimum in $-ln(P_{st}(n_{max}))$) and decays exponentially close to the absorbing boundary. At the higher temperature of $T=1080K$, post-critical clusters grow rapidly. For this reason, we sample $n_{max}$ with a higher frequency to obtain smoother data for $\tau_{MFPT}(n_{max})$ that captures the post-critical growth phase well.
\subsection{Comparison of free energy profiles at deep supercooling}\label{subsec:compare_results}
\begin{figure*}[htbp!]
\centering
\subfloat{\includegraphics[width=0.3\textwidth]{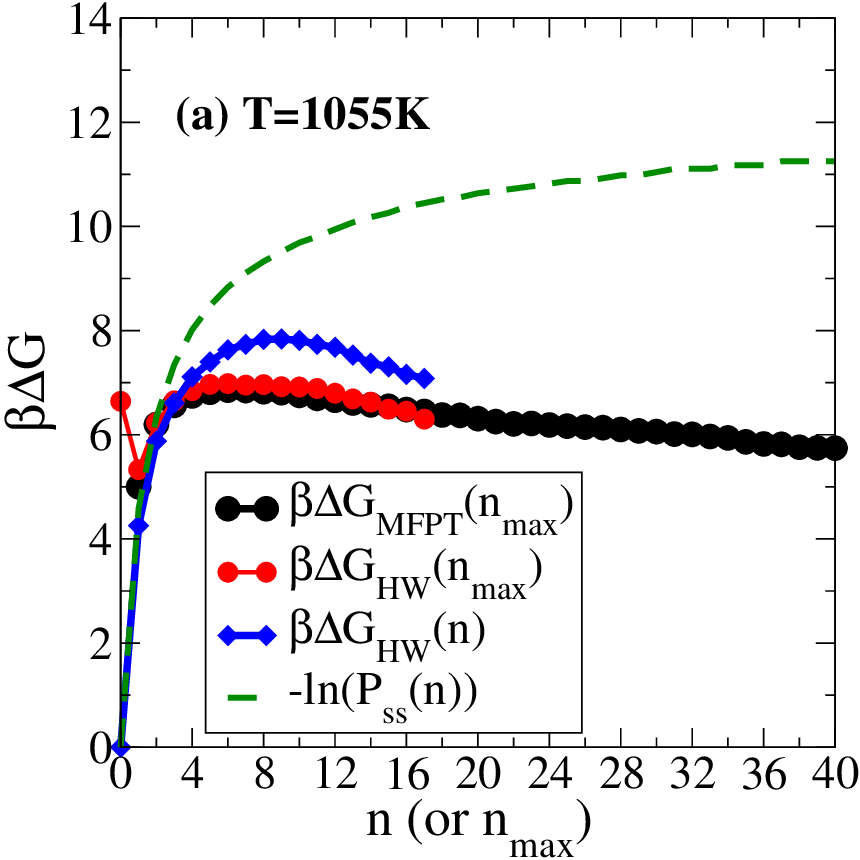}}
\subfloat{\includegraphics[width=0.3\textwidth]{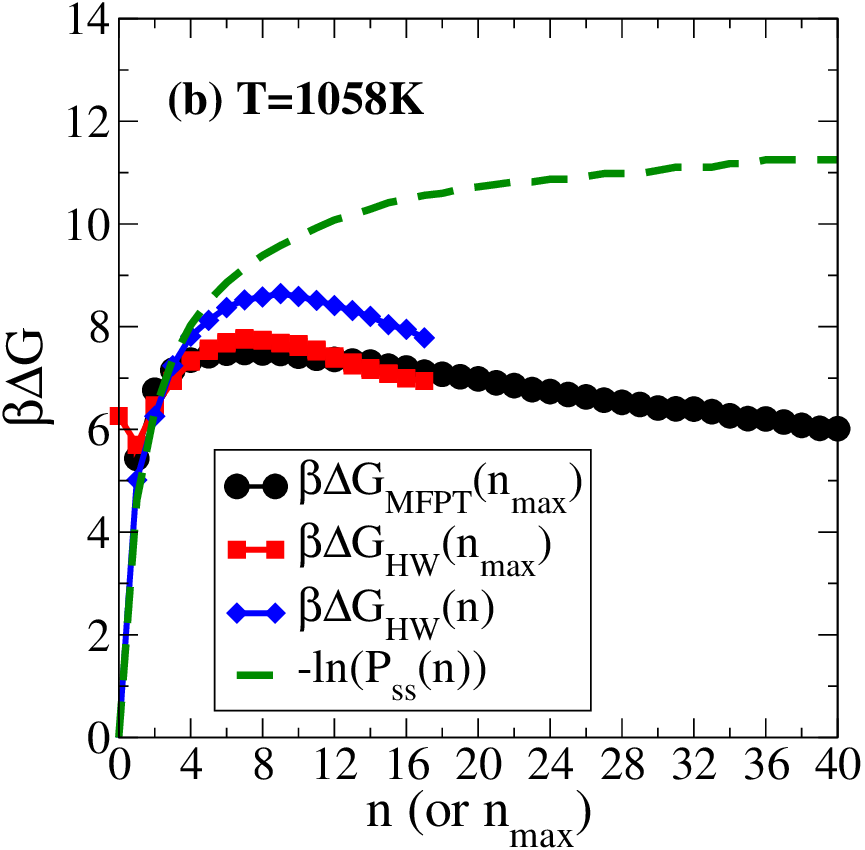}}
\subfloat{\includegraphics[width=0.3\textwidth]{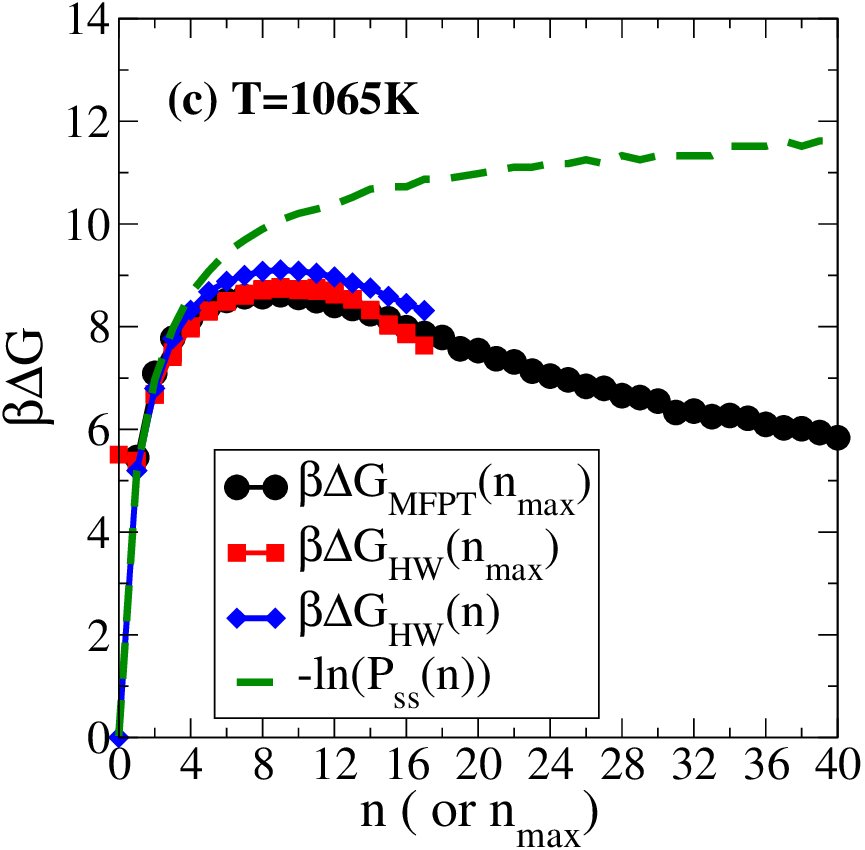}}

\subfloat{\includegraphics[width=0.3\textwidth]{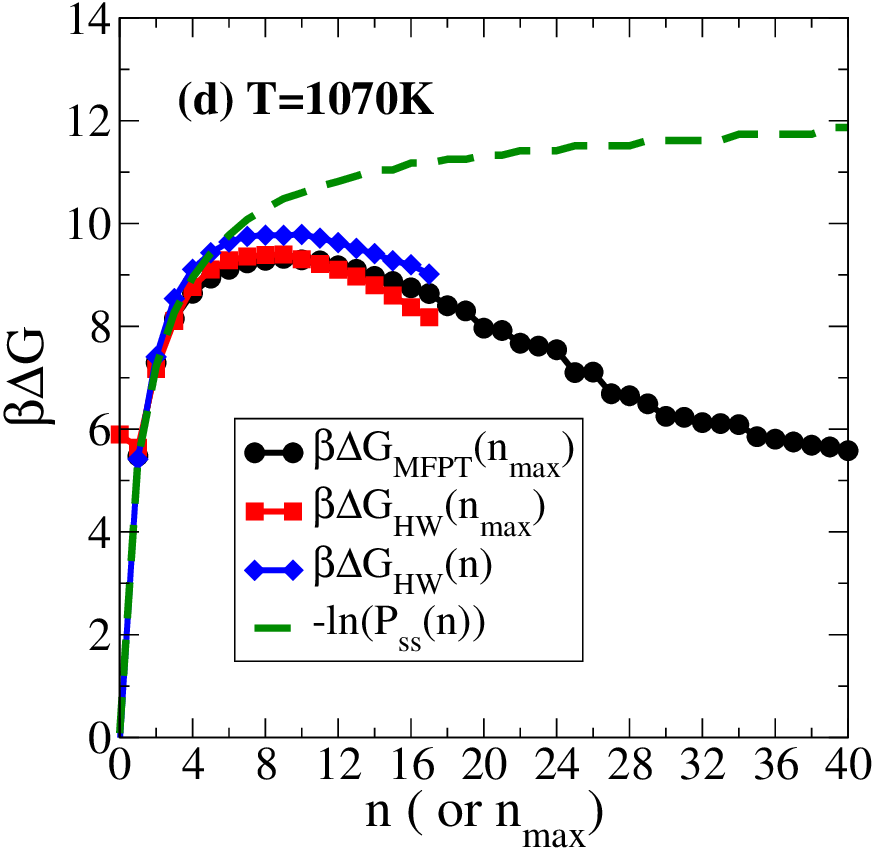}}
\subfloat{\includegraphics[width=0.3\textwidth]{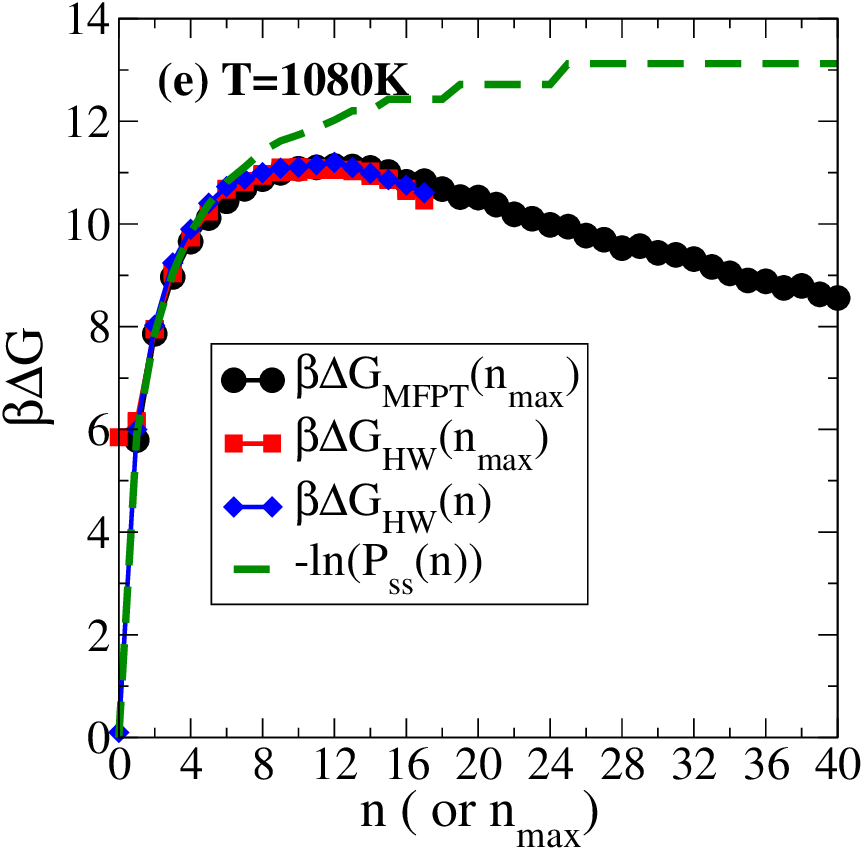}}
\caption{
Free energy profile obtained using the MFPT method and using umbrella sampling with a hard wall bias at (a) $T=1055K$, (b) $T=1058K$, (c) $T=1065K$, (d) $T=1070K$, (e) $T=1080K$. The procedure described in Appendix~\ref{subsec:nmax-v-n} is used to make a comparison between $\beta\Delta G(n_{max})$ and $\beta\Delta G(n)$. The free energy for the MFPT reconstruction is obtained from $600$ independent NPT MD runs of $N=512$ particles at $P= 0~GPa$. The free energy for the umbrella sampling runs is produced from simulations of $N=512$ particles at $P=0~GPa$. The small $n$ or $n_{max}$ behavior is obtained from $-ln(P_{ss}(n))$ as explained in the text, which is shown for comparison. 
}
\label{fig:fig-3}
\end{figure*}

\begin{figure}[htbp!]
\centering
\subfloat{\includegraphics[scale=0.5]{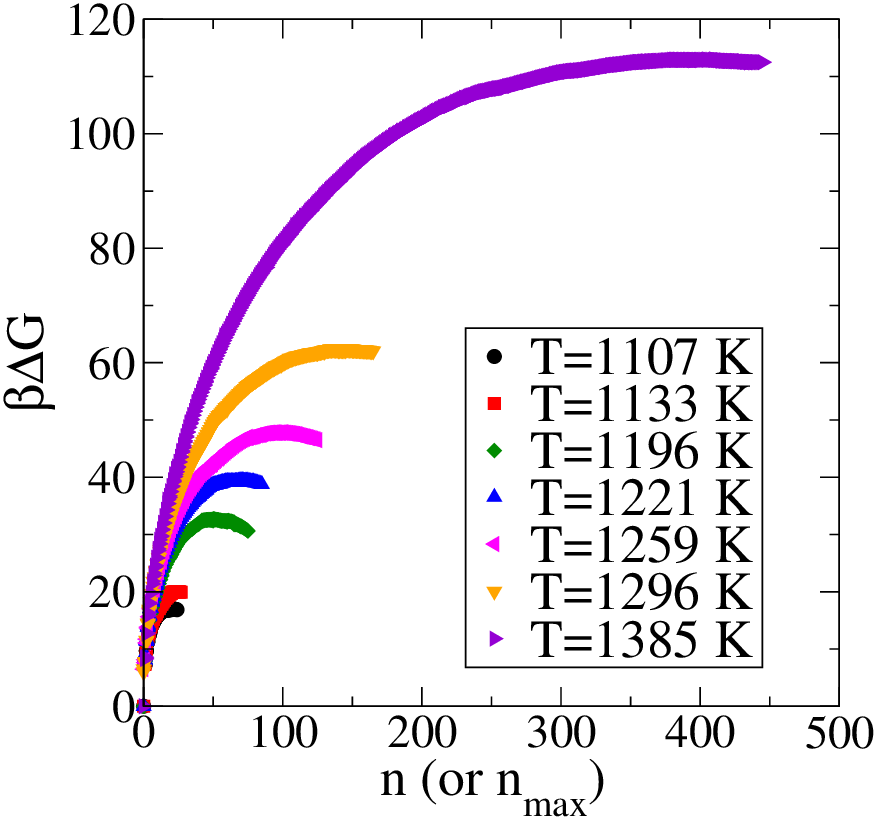}}
\caption{Free energy difference $\beta \Delta G$ against the nucleus size obtained
	from NPT umbrella sampling MC simulation at $P=0GPa$ with $N=4000$ at higher temperatures using  umbrella sampling runs with a harmonic bias and statistics of $n_{max}$ are gathered. Additional runs with a hard wall bias are performed, sampling $P(n)$, to improve statistics for small $n$ (or $n_{max}$}
        \label{fig:fig-4}
\end{figure}

The behaviour of supercooled liquid silicon is a matter of debate at deep supercooling, particularly in the vicinity of $T=1060K$ at $P=0~GPa$.
To address the question of whether crystallisation is spontaneous at these state points, the free energy barrier to the growth of crystalline clusters is calculated using the two methods described above. We find that a clear and significant barrier to the growth of the crystalline phase exists at each of the state points considered and that the two methods give results that are in agreement, shown in Fig.~\ref{fig:fig-3}.
Mendez-Villuendaz {\it et al}~\cite{mendez2007limit}, find that the largest cluster, $n_{max}$, is the appropriate order parameter to determine the thermodynamic stability of the parent phase based on stronger coupling between the nucleation kinetics and the free energy profile as a function of $n_{max}$ in the context of supercooled gold nanoclusters in the liquid phase. As also in other work,~\cite{bhimalapuram2007elucidating,chakrabarty2008chakrabarty}, a monotonically decreasing free energy as a function of $n_{max}$ is argued~\cite{mendez2007limit} to mark the loss of metastability of the liquid with respect to crystallisation. This conclusion is derived from the argument that $n_{max}$ is the order parameter that is best coupled to nucleation timescales.
However, the thermodynamic stability of the metastable liquid is determined by the free energy cost to the growth of any cluster of size $n$, $\beta\Delta G(n)$. In the present case, we point out that at all the state points we have considered, the free energy profile, $\beta\Delta G(n_{max})$, is not monotonically decreasing with $n_{max}$ and displays a clear barrier.
The procedure described in Appendix~\ref{subsec:nmax-v-n} is followed to produce a comparison between the Hard Wall bias umbrella sampling results and those from the kinetic reconstruction. 
At $T=1055K,~1058K$, we find that the difference between $P(n_{max})$ and $P(n)$ (or between the corresponding steady state probabilities for the MFPT results, $P_{st}(n_{max}))$ and $P_{ss}(n)$) persists to larger values of $n$ (or $n_{max}$) than at higher temperatures. It is worth noting that the comparison between $P_{st}(n_{max})$ and $P_{ss}(n)$ is only meaningful for $n$ or $n_{max})$ small enough that the steady state probabilities are good approximations to the equilibrium probabilities. When the difference between $P(n_{max})$ and $P(n)$ persists to larger values, the only meaningful comparison between results from the two methods, umbrella sampling and the MFPT method, are those where the order parameter is the same, namely, $n_{max}$. Notwithstanding the difficulty in making a satisfactory quantitative comparison with the free energy profiles obtained using the different methods at the lowest two temperatures, we close by pointing out two salient features of the results that are reported in Fig.~\ref{fig:fig-3}, which are central to the main focus of the present study: (i) At all temperatures studied, a clear and significant free energy barrier is present for crystal nucleation, and the different estimates, $\beta\Delta G_{MFPT}(n_{max})$, $\beta\Delta G_{HW}(n_{max})$ and $\beta\Delta G_{HW}(n)$ are in reasonable quantitative agreement. (ii) The free energy profiles $\beta\Delta G_{MFPT}(n_{max})$ and  $\beta\Delta G_{HW}(n_{max})$, obtained using the same order parameter $n_{max}$ are in very good quantitative agreement at all temperatures, including the lowest two temperatures at which their comparison with $\beta\Delta G_{HW}(n)$ is not very satisfactory.
At higher temperatures along the $P=0~GPa$ isobar (see Fig.~\ref{fig:fig-4}, where such ambiguities do not arise, we perform free energy calculations using umbrella sampling runs with a harmonic bias and study the effect of changes in the properties of the metastable liquid on the free energy barriers. The calculations are made starting from $T=1296 K$ ($\sim 23\%$ undercooling) to $T=1107K$ ($\sim 35\%$ undercooling).

\subsection{Free energy profiles along different isobars across the phase diagram}
\begin{figure}[h!]
\centering
\includegraphics[scale=0.5]{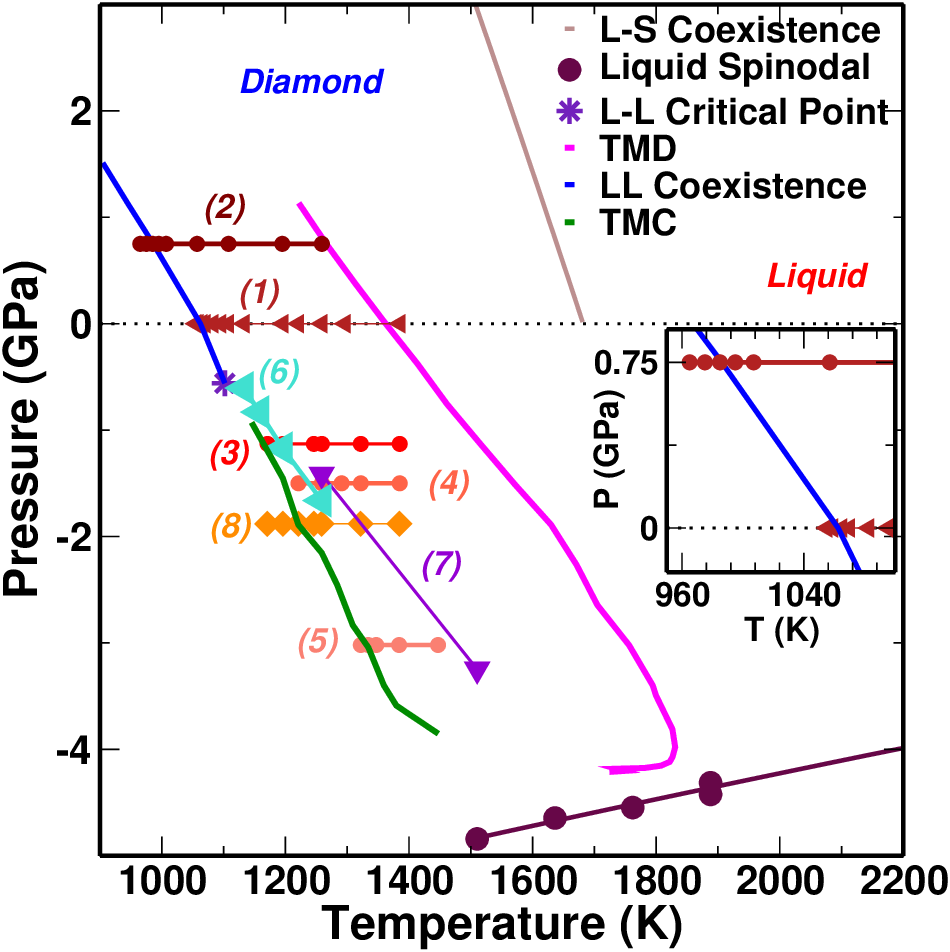}
\caption{The phase diagram of SW silicon showing the loci of interest and the isolines along which the free energy barrier to crystallisation is calculated. Each of the isolines is labelled, with the key as follows - {\bf (1)} $P=0~GPa$ isobar, {\bf (2)} $P=0.75~GPa$ isobar, {\bf (3)} $P=-1.13~GPa$ isobar, {\bf (4)} $P=-1.5~GPa$ isobar, {\bf (5)} $P=-3.02~GPa$ isobar, {\bf (6)} Line of constant coordination number $C_{NN}=4.66$  {\bf (7)} Line of constant isothermal compressibility $\kappa_T$, {\bf (8)} $P=-1.88~GPa$ isobar crossing the line of maximum isothermal compressibility. {\it Inset} Zoomed in to the temperatures along the $P=0~GPa$ and $P=0.75~GPa$ isobars at which the free energy calculation is performed, showing the estimated liquid-liquid coexistence line in blue.}
\label{fig:fig-5}
\end{figure}

\begin{figure*}[htpb!]
    \centering
    \subfloat{\includegraphics[scale=0.5]{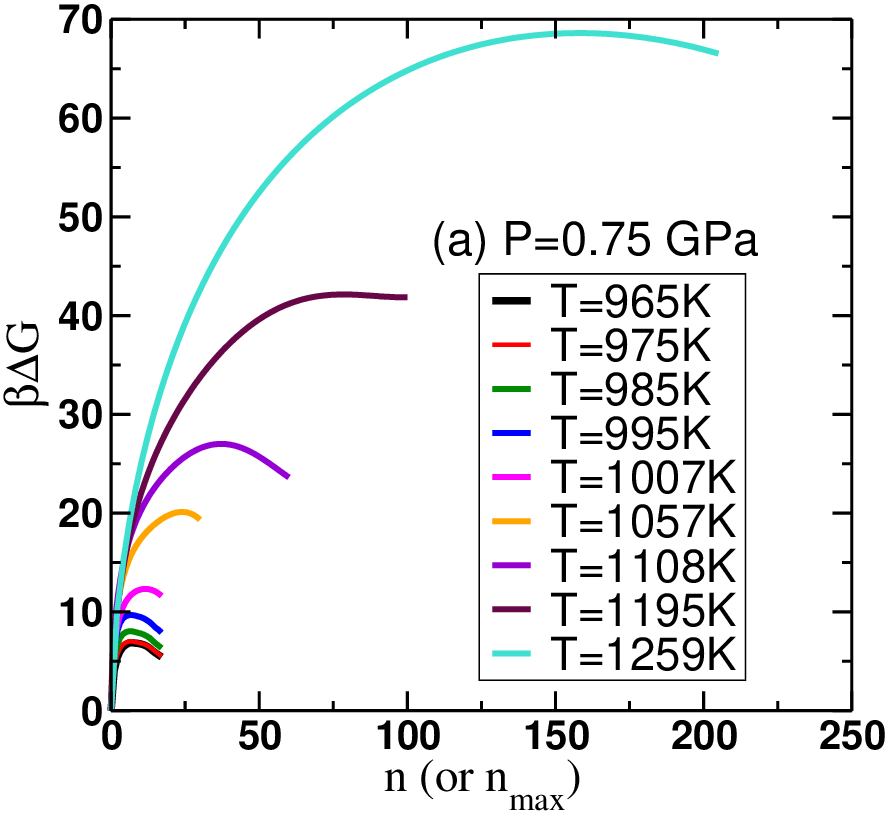}}
    \subfloat{\includegraphics[scale=0.5]{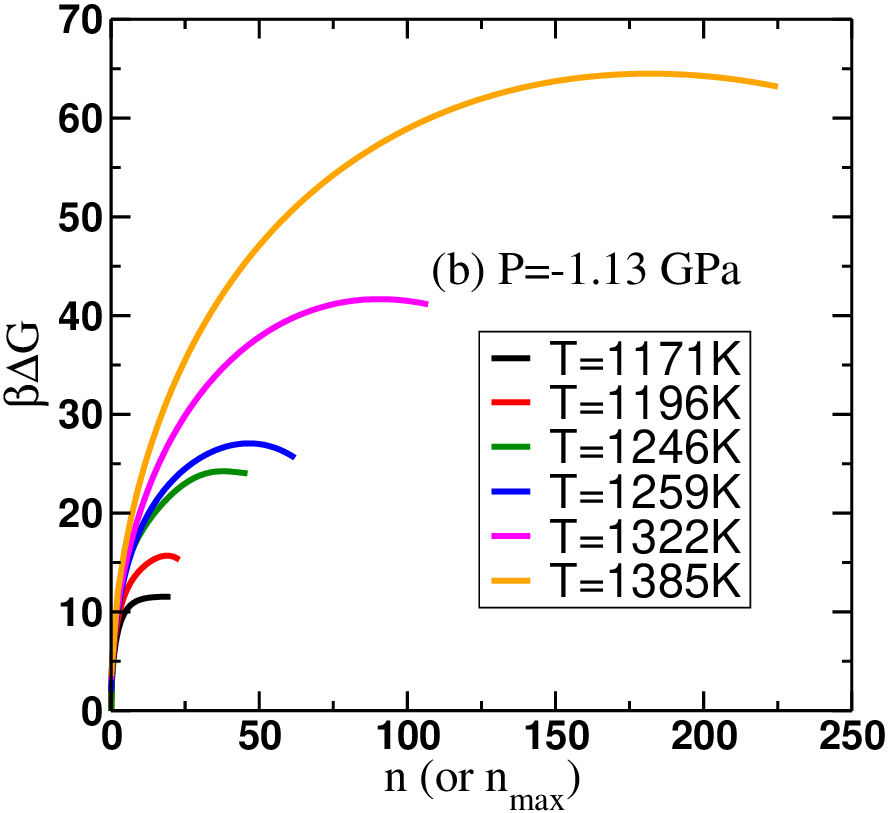}}

    \subfloat{\includegraphics[scale=0.5]{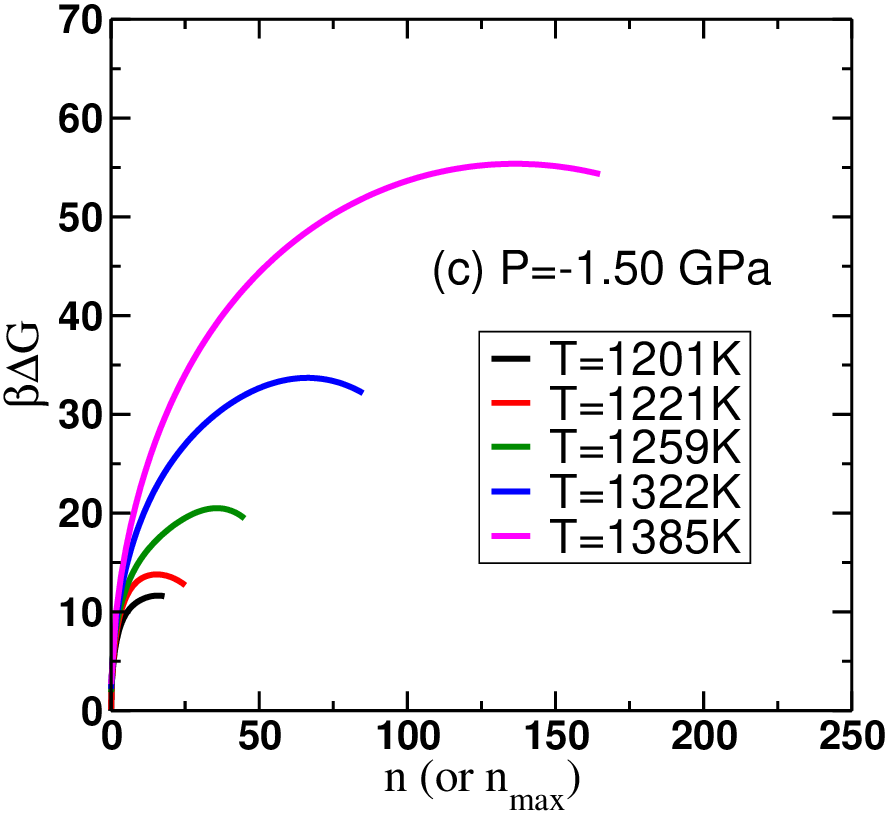}}
    \subfloat{\includegraphics[scale=0.5]{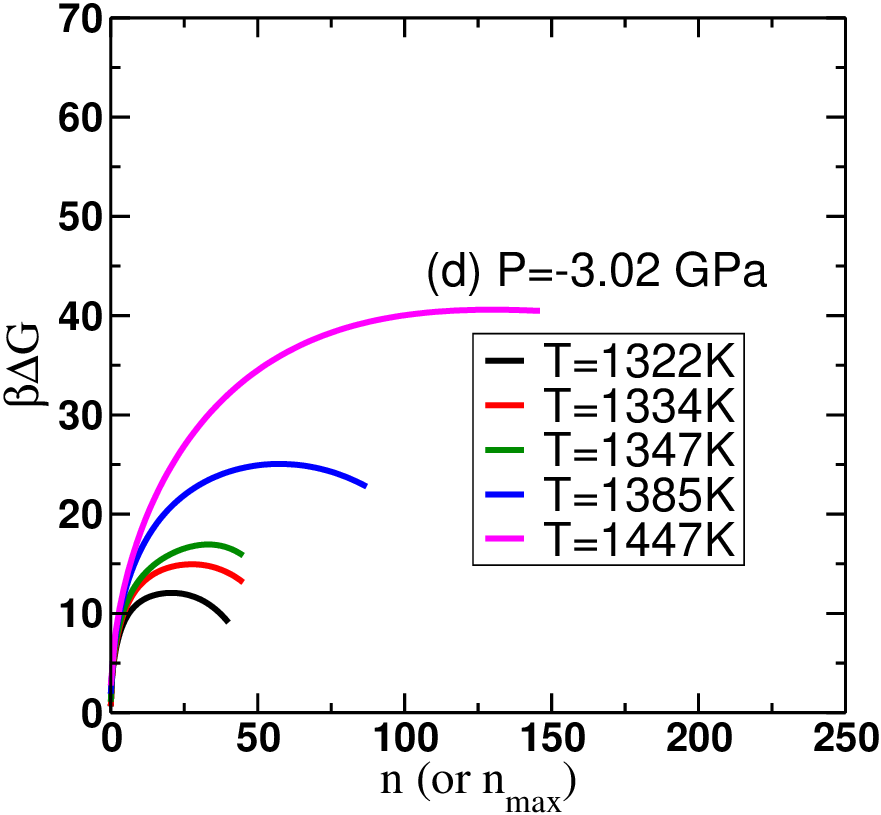}}
    \caption{Free energy difference $\Delta G(n_{max})$ against the nucleus size
 	    ($n_{max}$) obtained from NPT umbrella sampling MC simulation along the {\bf (a)} $P=0.75GPa$,{\bf (b)} $P=-1.13GPa$,{\bf (c)} $P=-1.51GPa$ and {\bf (d)} $P=-3.02GPa$ isobars with $N=4000$. For $T<1000K$ along the $P=0.75~GPa$ isobar, $\beta\Delta G(n)$ is calculated from $P(n)$ obtained from umbrella sampling runs with a hard wall bias. At each other temperature and pressure, umbrella sampling runs with a harmonic bias on $n_{max}$ is used, and statistics of $n_{max}$ are gathered. Additional runs with a hard wall bias are performed, sampling $P(n)$, to improve statistics for small $n$ (or $n_{max}$.}
    \label{fig:fig-6}
\end{figure*}

In the next set of results, the free energy profiles at different state points, indicated in Fig.~\ref{fig:fig-5}, are calculated.
In Fig.~\ref{fig:fig-6}, we show the $\beta \Delta G$ computed across a range of temperatures for $P=0.75 GPa$, $-1.13GPa$, $-1.51 GPa$ and $-3.02 GPa$ respectively. With the exception of temperatures below $T=1000K$ in Fig.~\ref{fig:fig-6} {\bf (a)}, these state points correspond to the HDL region of the phase diagram as understood from the results of Ref.~\onlinecite{vasisht2011liquid}. The free energy profiles at  $P= - 1.88 GPa$, for state points across the Widom line, are discussed separately below. 

\subsection{Free Energy profiles and Compressibility}\label{subsec:Kt_results}

The next question of interest is how free energy barrier changes as the reported critical point is approached from other isolines, namely the line of constant coordination number, where the isothermal compressibility increases as we approach the critical point, and the line of constant isothermal compressibility, where the coordination number changes. The state points of both sets of data are chosen such that they are roughly parallel to both the line of compressibility maxima (the Widom line) and the line of maximum density (see Fig.~\ref{fig:fig-5}).
We first describe the results for state points of varying compressibiliity, but keeping the coordination number fixed. 

\begin{figure}[htp!]
\centering
\includegraphics[scale=0.5]{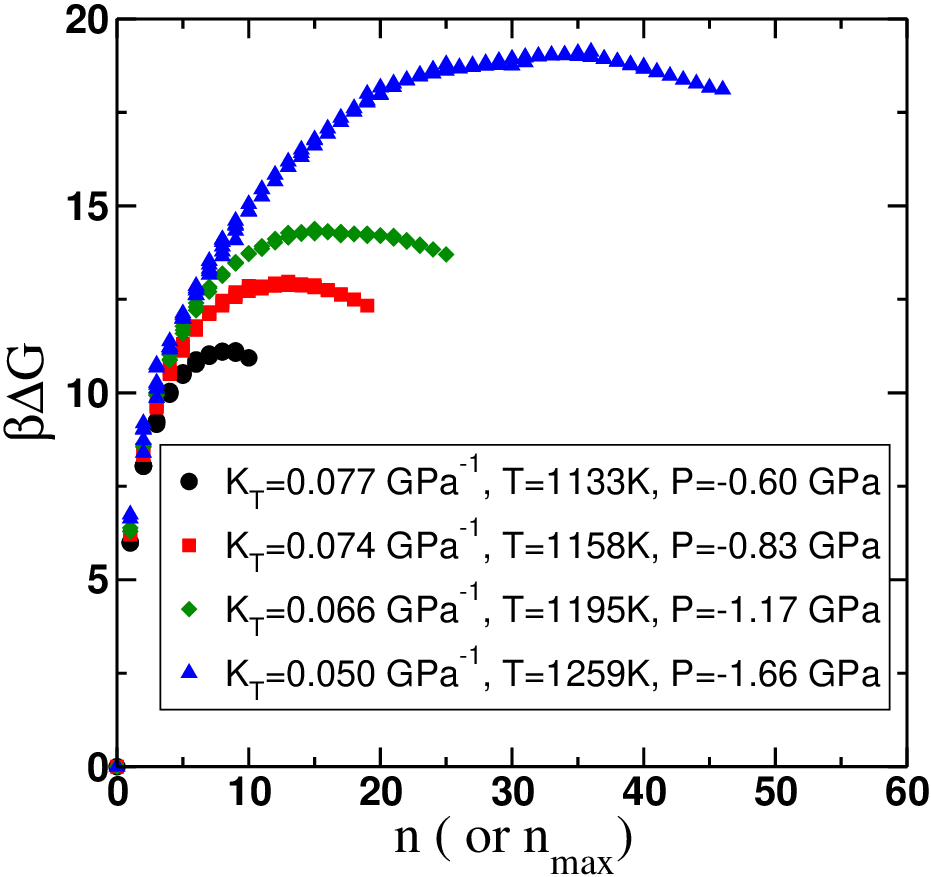}
\caption{Free energy difference $\beta \Delta G$ against the nucleus size
	    obtained from NPT umbrella sampling MC simulations with a system size of $N=4000$. At all the state points,
	    the coordination number is the same ($C_{nn}=4.66$), but the compressibility decreases monotonically with temperature and pressure. For each of the curves, umbrella sampling runs are performed with a harmonic bias on $n_{max}$, and statistics of $n_{max}$ are gathered. Additional runs with a hard wall bias are performed, sampling $P(n)$, to improve statistics for small $n$ (or $n_{max}$).}
\label{fig:fig-7}
\end{figure}

A number of studies in the literature have highlighted the role of enhanced fluctuations in the metastable liquid in reducing the free energy barrier to crystal nucleation~\cite{ten1997enhancement,russo2012microscopic,kurita2019drastic}.
In these cases, the enhanced fluctuations are brought about by proximity to a fluid-fluid phase transition or fluid-fluid critical point. To understand the effect of fluctuations on the free energy barriers for supercooled liquid silicon, we construct the free energy profile along a locus of constant coordination number, or degree of tetrahedrality, of the metastable liquid with varying compressibility.
We have chosen state points such that the coordination number remains constant ($C_{nn}=4.66$) as the compressibility increases.
The isothermal compressibility of the liquid changes along the line of state points considered while the overall tetrahedral character of the liquid is fixed; we can analyse the effect of density fluctuations on the free energy barrier. We find that all these state points sit parallel to the line of compressibility maxima reported in Ref.~\onlinecite{vasisht2011liquid}, also known as the Widom line.
In Fig. \ref{fig:fig-7}
we observe that as we approach the critical point, the compressibility increases and the free energy barrier decreases to around $10 k_B T$.
The critical nucleus size changes from $35$ atoms to less than $10$ atoms. While it is difficult to determine from this analysis whether the decrease in the work required for samples to crystallise is determined by the fluctuations in density or by possibly associated fluctuations in bond order, suggested in earlier work on hard sphere crystallisation~\cite{russo2012microscopic}, we highlight that larger fluctuations appear to destabilise the metastable liquid with respect to crystallisation as also seen in previous work.
Further, we reiterate that the state points chosen fall on a locus of constant coordination number and the average degree of tetrahedrality in the liquid is expected to be the same.
\begin{figure}[htp!]
\centering
\includegraphics[scale=0.5]{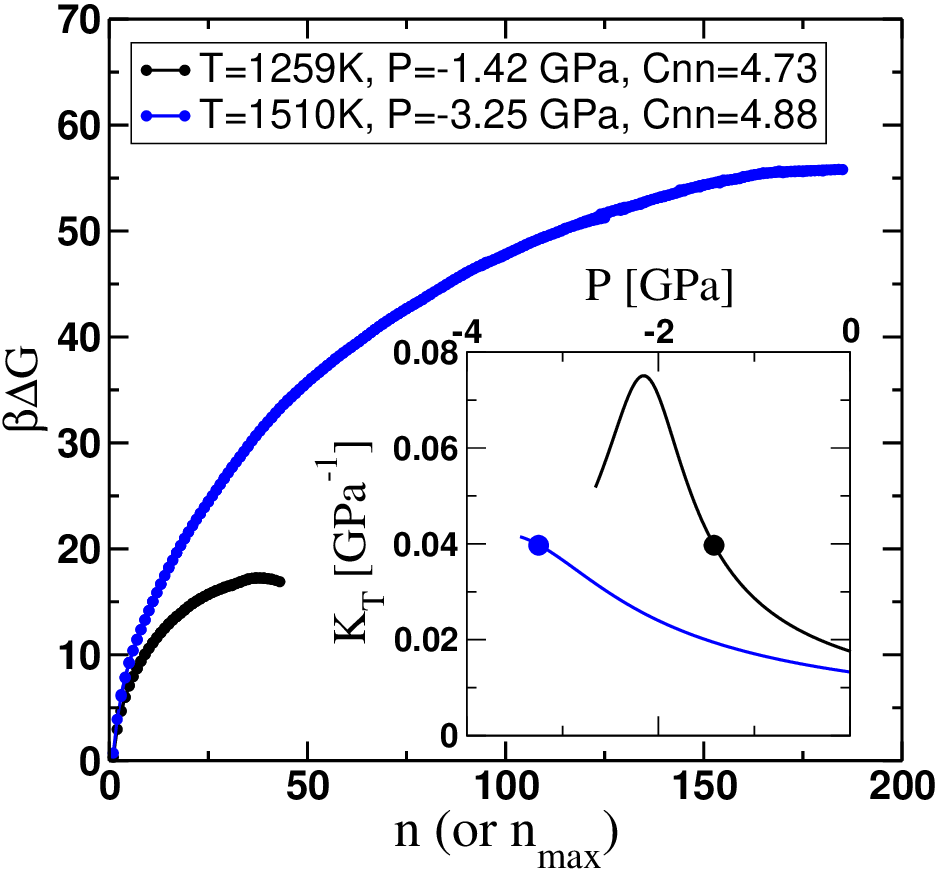}
\caption{Free energy difference $\beta \Delta G$ against the nucleus size
	    obtained from NPT umbrella sampling MC simulations with $N=4000$. At both the state points,
	    the compressibility is fixed ($K_{T}=1.5$ reduced units), but the coordination
	    number changes by $3\%$. Umbrella sampling runs are performed with a harmonic bias on $n_{max}$, and statistics of $n_{max}$ are gathered. Additional runs with a hard wall bias are performed, sampling $P(n)$, to improve statistics for small $n$ (or $n_{max}$).}
	    \label{fig:fig-8}
\end{figure}

\subsection{Free Energy profiles and Coordination Number}\label{subsec:CNN_results}
We next look at the effect of local coordination number on the free energy barrier keeping
the compressibility fixed.
Keeping the compressibility fixed we try to find the effect of coordination number on $\beta\Delta G$ of the system. We chose the compressibility value such that the difference in the coordination number is largest between the two state points accessible at that compressibility.  In Fig.~\ref{fig:fig-8} (inset) we show the dependence of compressibility on the pressure for two isotherms $T=1259 K$ and $T =1510 K$.
The symbols in the inset represent the chosen state points, which have equal compressibility and a difference in coordination number of $3\%$. In Fig. \ref{fig:fig-8} (main panel) we show the corresponding change in the free energy, wherein we find a dramatic change in the free energy barrier and critical nucleus size as we approach the smaller coordination number.
The data suggest that an increase in the tetrahedral ordering in the parent liquid can significantly alter the characteristics of the crystallisation transition. However, it is difficult to conclude what drives the transition the most.
Given the stark difference in the barrier to crystal nucleation, understanding the observed free energy barrier remains an interesting question.
\subsection{Free Energy profiles at $P=-1.88~GPa$, across the Widom line}\label{subsec:Widom_results}

\begin{figure}[htbp!]
\centering
\subfloat{\includegraphics[scale=0.52]{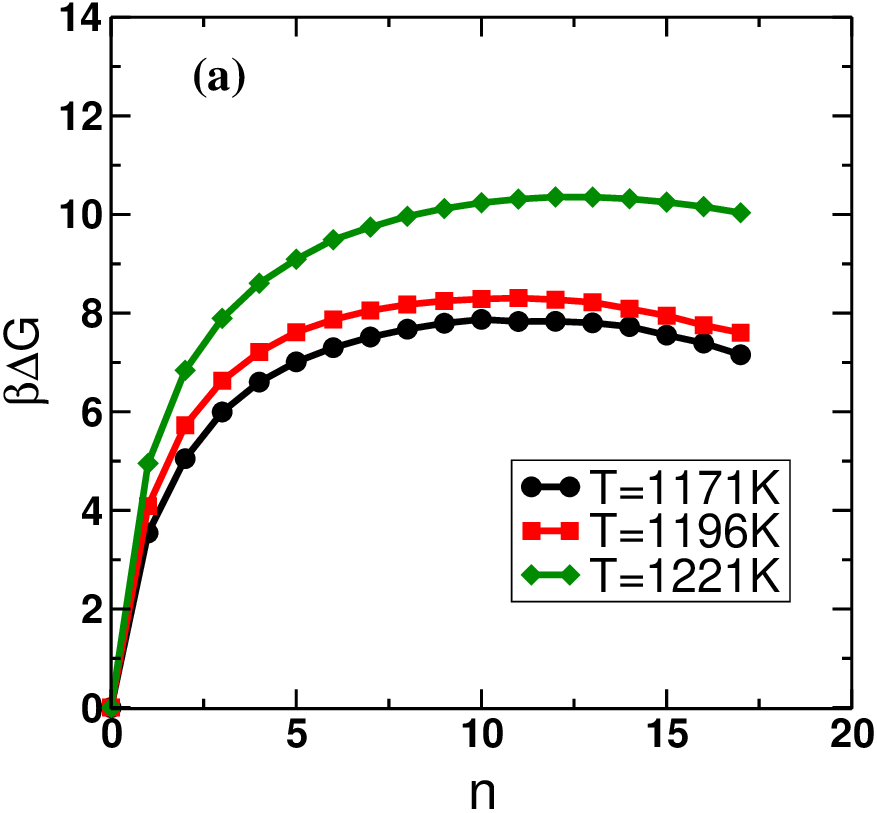}}

\subfloat{\includegraphics[scale=0.52]{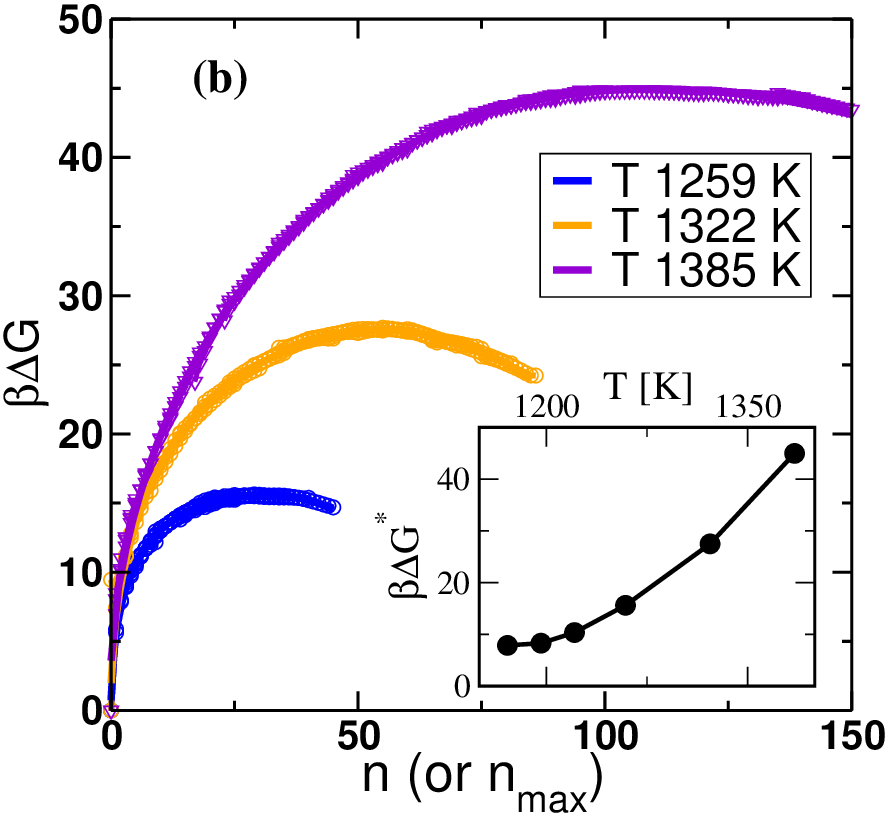}}
\caption{Free energy difference $\beta \Delta G$ against the nucleus size obtained
	from NPT umbrella sampling MC simulation at $P=-1.88GPa$. (a) Low temperature side of Widom line, $N=512$ (b) High temperature side of Widom line, $N=4000$. For the low temperature side in {\bf (a)}, the full cluster size distribution, $P(n)$, is obtained from runs with a hard wall bias on $n_{max}$ and used to construct the free energy curves. For the high temperature side, {\bf (b)}, umbrella sampling runs are performed with a harmonic bias on $n_{max}$, and statistics of $n_{max}$ are gathered. Additional runs with a hard wall bias are performed, sampling $P(n)$, to improve statistics for small $n$ (or $n_{max}$). The inset shows the temperature dependence of the free energy barrier height, $\beta\Delta G^{*}$ and the critical nucleus size $n^{*}$.}
        \label{fig:fig-9}
\end{figure}

We next evaluate the change in $\beta\Delta G$ across the Widom line for the $P=-1.88GPa$ isobar. The line of compressiblity maxima, called the Widom line, that extends beyond the liquid-liquid critical point in water and related systems, has been the focus of several studies~\cite{xu2005relation,kumar2005static,abascal2010widom,kesselring2013finite,lascaris2013response}. In these studies, sharp (if continuous) changes in various properties have been reported across the Widom line. We investigate whether crossing this line at constant pressures below the critical pressure reveals any indication of a marked change in the nucleation barriers
The pressure is fixed at $P=-1.88GPa$, a value lower than the reported critical point, and the temperature varied from $T=1385K$ to $T=1171K$. The compressibility maximum at $P=-1.88~GPa$ is at $T\sim1230K$. The free energy barrier is found to decrease monotonically with temperature in Fig.~\ref{fig:fig-9}. On the low temperature side of the reported Widom line, the free energy barrier changes by $2 k_B T$ for a $50K$ change in temperature while on the high temperature side, we find that for a similar change in temperature, the free energy barrier changes by $>10 k_B T$. Thus, our results indicate that indeed, a change in the temperature dependence of nucleation barriers occurs upon crossing the Widom line. Considering the critical nucleus size, $n^{*}$, we find a more striking change, with the critical nucleus size becoming nearly constant below the Widom line. While such a change in temperature dependence is of interest, the presence of a free energy barrier to crystallisation exists at state points above and below the Widom line, and points to the liquid retaining metastability at all these state points. 

\subsection{Dependence of barrier height and critical size on temperature}

\begin{figure}[htbp!]
\centering
\subfloat{\includegraphics[scale=0.52]{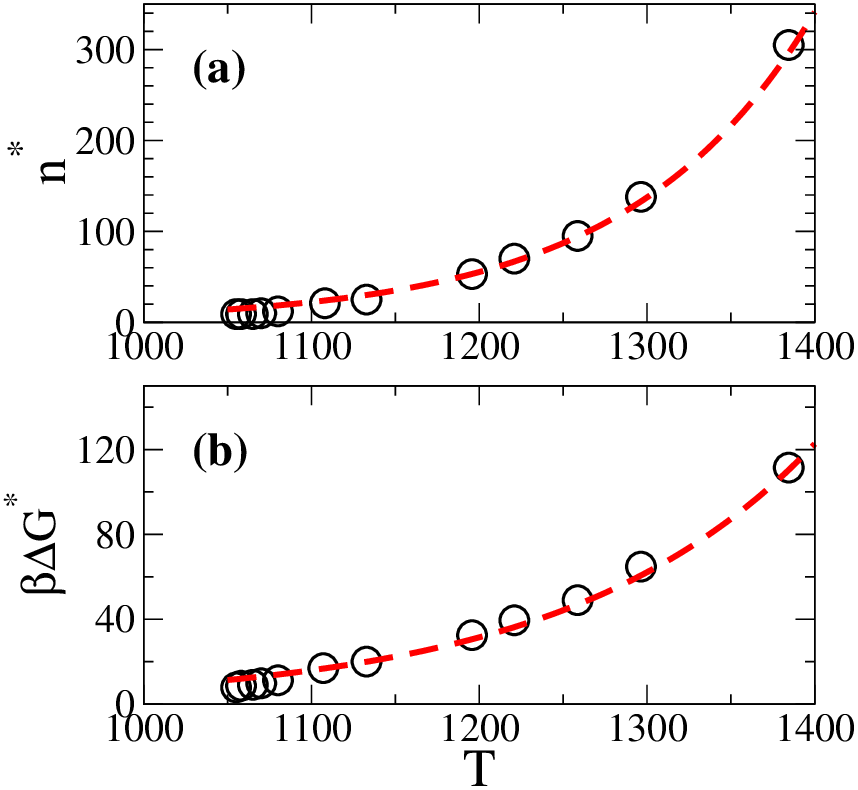}}

\subfloat{\includegraphics[scale=0.5]{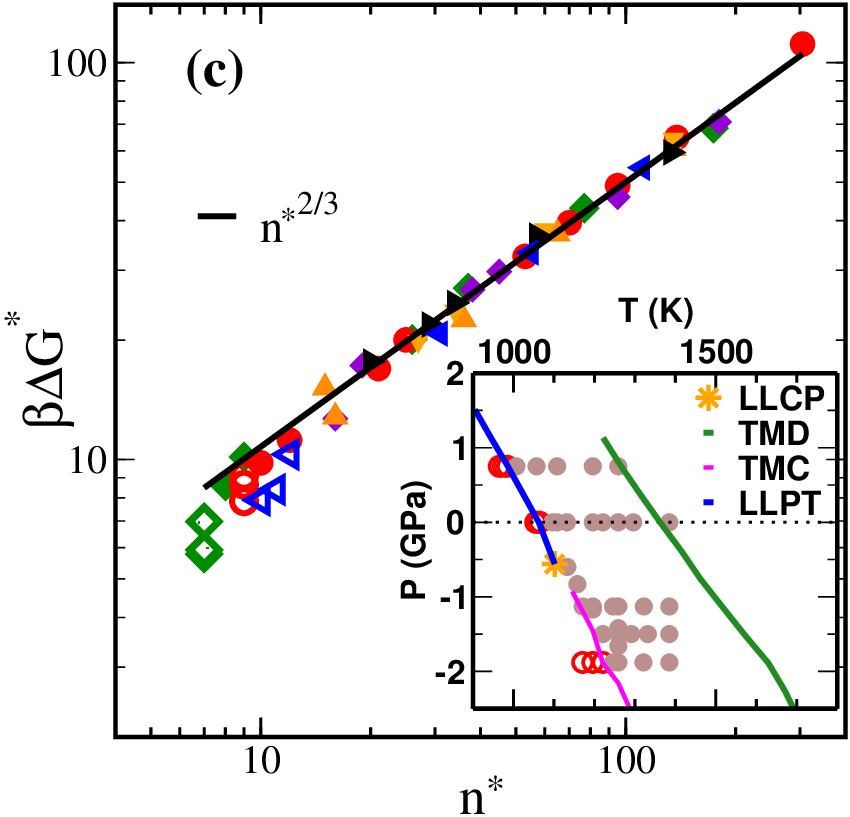}}
\caption{{\bf (a)} The critical cluster size $n^{*}$ and {\bf (b)} the height of the free energy barrier, $\beta\Delta G^{*}$ as a function of temperature at $P=0~GPa$. The dashed lines are guides to the eye. {\bf (c) } A parametric plot of the barrier height, $\beta \Delta G^{*}$ and the corresponding critical cluster size, $n^{*}$ obtained from different isobars. The solid line corresponds to the $n^{2/3}$ dependence of the free energy barrier expected according to CNT. Data points found to deviate from the scaling at low $n^{*}$ are shown as open symbols in the main panel and the corresponding temperatures and pressures are marked as open red circles in the inset. The solid brown points in the inset are the state points for which the free energy barrier varies as  $n^{2/3}$, which are shown as solid symbols in the main panel.}
        \label{fig:fig-10}
\end{figure}

Trends in the critical cluster size $n^{*}$ and the barrier height, $\beta \Delta G^{*}$, as a function of temperature along the $P=0~GPa$ isobar are shown in Fig.~\ref{fig:fig-10} {\bf (a)} and {\bf (b)}. 
Fig.~\ref{fig:fig-10} {\bf (c)} contains a parametric plot of the barrier height and the corresponding cluster size. Interestingly, one finds that a $n^{2/3}$ scaling of the barrier height fits the data well at state points where $n^{*}$ is large. This is in accordance with the CNT prediction. At deep supercooling where the critical cluster is small and poorly approximated to a sphere, the predictions from CNT are not expected to be obeyed given that a number of the assumptions made in CNT are not satisfied when $n^{*}$ is small. Interestingly, we find that all the state points which show deviations from the CNT prediction fall on the lower temperature side of the liquid-liquid phase transition or the Widom line estimated in Ref. \onlinecite{vasisht2011liquid}. 

\subsection{Effect of changing the ensemble of starting configurations at $T=1055K$}

\begin{figure}[htp!]
    \centering
    \includegraphics[scale=0.5]{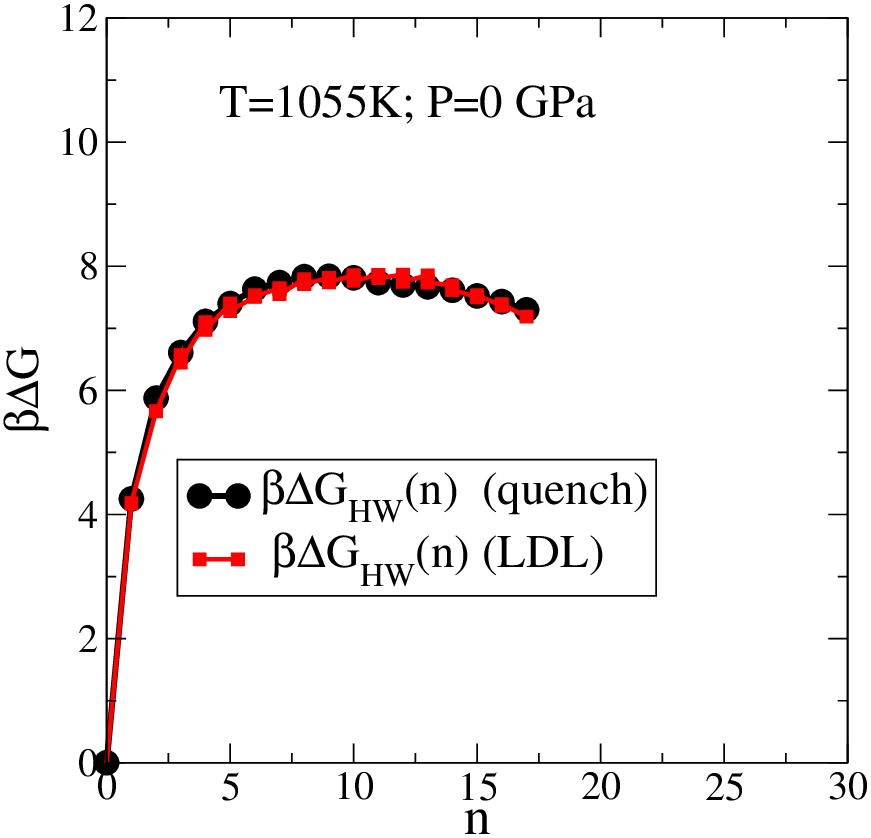}
    \caption{Comparison of the free energy profiles from two sets of umbrella sampling runs with the hard wall bias. Simulations are performed at $T=1055K,~P=0~GPa$ with a system size $N=512$. Data points labelled "quench" are obtained starting from a set of randomly placed particles in a box such that the density is $2.48~gcc^{-1}$. Data points labelled "LDL" are obtained starting from a set of configurations selected from MD runs at $T=1055K,~P=0~GPa$ with low density and low degree of crystallinity as discussed in the text. The full cluster size distribution, $P(n)$, is obtained from runs with a hard wall bias on $n_{max}$ and used to construct the free energy curves.}
    \label{fig:fig-11}
\end{figure}

We compare the free energy curves produced when the ensemble of initial conditions is changed from the disordered liquid considered earlier to a liquid more typical of $T=1055K$, noting the significant difference in the characteristics of the two at this temperature. One expects that if the sampling along other order parameters can be assumed to be complete, regardless of the set of starting configurations, then the two sets of results should be exactly the same. This is seen in the case of the umbrella sampling runs with the hard wall bias, shown in Fig.~\ref{fig:fig-11}. We compare results when the starting configurations are of randomly place particles in a box corresponding to density $2.48~gcc^{-1}$ to those where the starting set of configurations are selected from MD runs at $T=1055K,~P=0~GPa$ satisfying the following criteria: (a) $\rho~\leq~2.37~gcc^{-1}$, (b) $n_{max}~\leq~5$, and (c) $n_{tot}~\leq~10$ where $n_{tot}$ is the total number of crystalline atoms. The resulting free energy profiles display no dependence on the initial ensemble of configurations. 


\section{Discussion} \label{sec:discussion}

In summary, we have investigated crystal nucleation barriers in Stillinger-Weber silicon for a wide range of state points, employing two distinct methods, namely umbrella sampling and the reconstruction of the free energy barriers through the computation of mean first passage times from unconstrained molecular dynamics simulations. In particular, we focus on state points close to the liquid-liquid transition that has previously been studied, although in the present work we do not directly address the liquid-liquid transition itself. Instead, we focus on the question of whether a free energy barrier to crystal nucleation exists at the relevant state points, since it has been suggested in some previous works that no such barrier exists and crystal nucleation occurs spontaneously \cite{limmer2011putative,limmer2013putative,ricci2019computational}. Based on the two independent methods of estimating free energy barriers mentioned above, we consistently find that at all state points we investigate, finite free energy barriers to crystallisation are present.
Thus, our results confirm that in small systems such as are often used in simulations, a barrier to crystallisation exists, and the supercooled liquid therefore has a well-defined metastable state. The resulting nucleation rates for $N = 512$ (Fig.~\ref{fig:fig-2}) are of the order of $2.5\times~10^7 s^{-1}$. The corresponding nucleation rates for macroscopic or even nanoscopic droplets (e.g., sub-micron droplets) would be very large, 
and the liquid would be too short-lived to be probed under normal experimental conditions, and would require well-designed, ultrafast, measurements to detect \cite{beye2010liquid}.
In addition to the low temperature state points at zero pressure that we focus on primarily, we compute the free energy barriers across a wide range of temperatures and pressures. We show that an increase in compressibility at fixed coordination number as well as a decrease in coordination number towards the tetrahedral value of $4$ at fixed compressibility, lead to a strong decrease in the free energy barriers. We show that crossing the Widom line at constant pressure leads to a change in the temperature dependence of the free energy barrier and the critical nucleus size -- both become slower functions of temperature -- indicating a change in the character of the liquid across the Widom line. 

Finally, we compare the dependence of the free energy barrier height, $\beta \Delta G^{*}$ on the size of the critical nucleus, $n^{*}$, with the CNT prediction that $\beta \Delta G^{*} \sim n^{*^{2/3}}$. Remarkably, we find that the CNT prediction is satisfied for the high temperature and pressure state points, that lie above the boundary defined by the liquid-liquid transition line and the Widom line taken together, as estimated in Ref. \cite{vasisht2011liquid}, and one observes deviations from the CNT prediction for state points across this boundary. Clearly, a change in character of the liquid takes place across this boundary. 

Our results thus clearly establish that finite barriers to crystal nucleation exist at state points across which a liquid-liquid transition have been argued to exist for Stillinger-Weber silicon by some previous works \cite{vasisht2011liquid} and where the metastability if the liquid has been questioned in others \cite{limmer2011putative,limmer2013putative,ricci2019computational}. They also point to changes in the nature of these barriers across state points which have been identified previously as corresponding to the liquid-liquid transition or the Widom line. These results do not directly address the existence of the liquid-liquid transition itself, but establish the necessary condition for questions about such a possibility to be meaningfully investigated. Ascertaining the existence of a liquid-liquid transition in a manner that satisfactorily addresses doubts that have been raised in previous work is the subject of future investigation. 

\section*{Acknowledgements}
The authors gratefully acknowledge the Thematic Unit of Excellence on Computational Materials Science, and the National Supercomputing Mission facility (Param Yukti) at the Jawaharlal Nehru Center for Advanced Scientific Research for computational resources.  SS acknowledges support through the JC Bose Fellowship  (JBR/2020/000015) SERB, DST (India). PGD gratefully acknowledges support from the National Science Foundation (grant CHE-1856704). The authors acknowledge with gratitude the intense interactions with C. A. Angell regarding the subject of the manuscript, and his insightful suggestions and ideas. Sebastiano Bernini is acknowledged for discussions and help in the early stages of this work. 


\section*{Data Availability Statement}
The data that support the findings of this study are available from the corresponding author upon reasonable request.
\FloatBarrier
\renewcommand{\thetable}{A\arabic{table}}  
\renewcommand{\thefigure}{A\arabic{figure}}
 \renewcommand{\theequation}{A\arabic{equation}}
 \setcounter{figure}{0}
 \setcounter{table}{0}
 \setcounter{equation}{0}
\appendix
\section{Detailed model and methods}\label{sec:Appendix_Model_and_Methods}
\subsection{The Stillinger-Weber potential} \label{subsec:potential}
The Stillinger-Weber potential, employed here, is the most widely used classical potential of silicon. It consists of a two-body term and a three-body term, $U_2$ and $U_3$, respectively.~\cite{stillinger1985computer}
\begin{equation}
U_{SW} = \sum_{j>1}^N U_2(r_{ij}) + \sum_{i<j<k}^N U_3({\bf r_i,r_j,r_k})
\end{equation}
The vectors, ${\bf r}_i$,${\bf r}_j$,${\bf {r}_k}$, are position vectors for atoms $i,j,k$ and $r_{ij}$ is the distance between the $i^{th}$ and $j^{th}$ atoms. N is the total number of atoms in the system.
\begin{equation}
U_2(r_{ij})=
\begin{cases}
\quad \epsilon A \left( \frac{B}{r^4_{ij}} -1 \right)e^{\frac{1}{r_{ij} - r_c} } \quad &if \quad r<r_c\\
\quad 0 \quad &if \quad r \ge r_c\\
\end{cases}
\end{equation}
The three-body interaction term is defined by
\begin{equation}
\begin{split}
U_3({\bf r_i,r_j,r_k}) = h(r_{ij},r_{ik},\theta_{jik})+h(r_{ij},r_{jk},\theta_{ijk})+\\h(r_{ik},r_{jk},\theta_{ikj})
\end{split}
\end{equation}
In turn, 
\begin{widetext}
\begin{equation}
h(r_{ij},r_{ik},\theta_{jik})=
\begin{cases}
\quad \epsilon ~\lambda \left[ cos\theta_{jik} + \alpha \right]^2 e^{\frac{\gamma}{r_{ij}-r_c} + \frac{\gamma}{r_{ik}-r_c}} \quad &if \quad r_{ij},r_{ik} < r_c\\
\quad  0 &if \quad r_{ij}\quad or \quad r_{ik} \ge r_c\\
\end{cases}
\end{equation}
\end{widetext}
The constants used in the equations above are listed in the table below:
\begin{center}
\begin{tabular}{ c c c c c c c}
\hline \hline 
Symbol & $A$ & $B$ & $r_c$ & $\lambda$ & $\alpha$ & $\gamma$ \\
\hline
Value & $7.04955$ & $0.60222$ & $1.80$ & $21.0$ & $1/3$ & $1.20$ \\
\hline \hline
\end{tabular}
\end{center}
\subsection{Order parameters} \label{subsec:OP}
\begin{figure}[h!]
    \centering
    \includegraphics[scale=0.45]{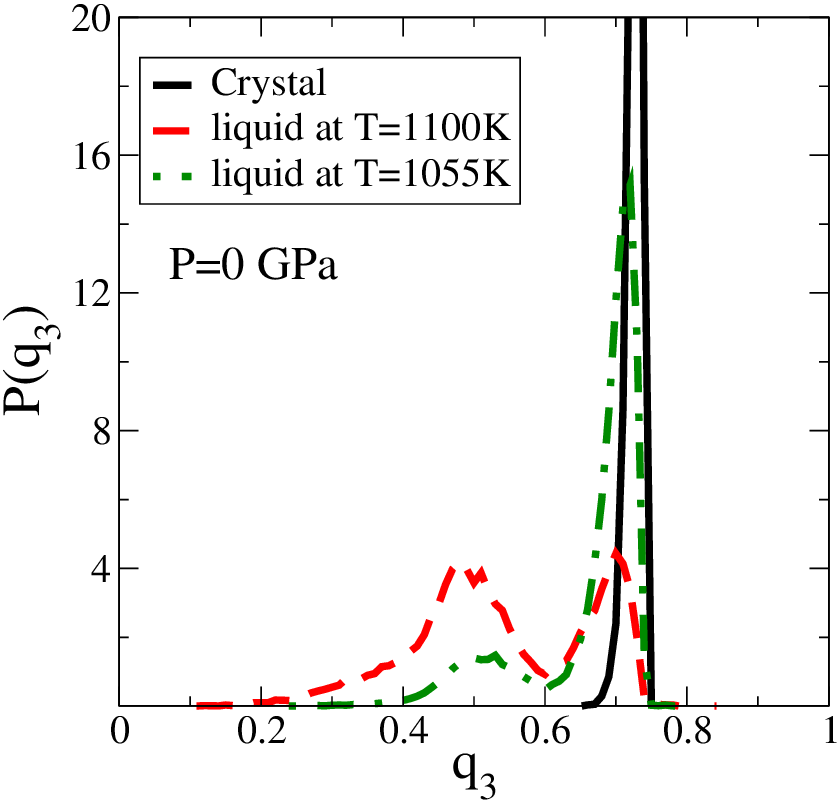}
    \caption{Distribution of $q_3$ for different types of configurations, high density liquid, crystal at $T=1260K$ and a non-crystallised, low density liquid (LDL) configuration.}
    \label{fig:fig-a1}
\end{figure}

\begin{figure}[h!]
    \centering
    \includegraphics[scale=0.45]{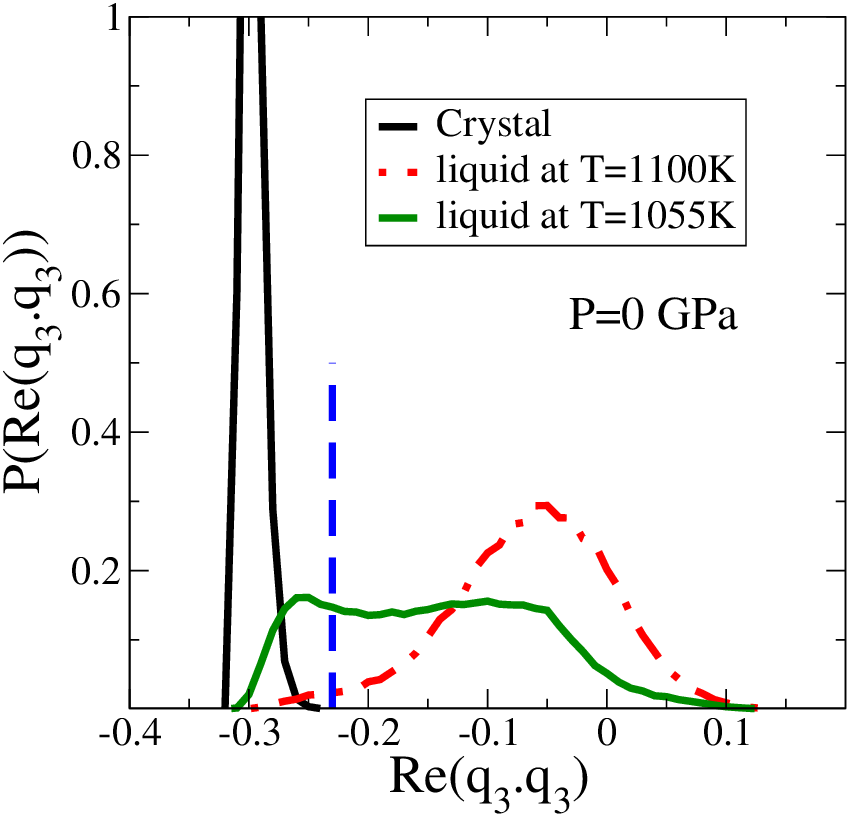}
    \caption{Distribution of $Re(q_3(i).q_3(j))$ for different types of configurations, high density liquid, crystal at $T=1260K$ and a non-crystallised, low density liquid (LDL) configuration. The blue vertical line at $-0.23$ marks the cutoff defining solid particles.}
    \label{fig:fig-a2}
\end{figure}

\begin{figure}[h!]
    \centering
    \includegraphics[scale=0.45]{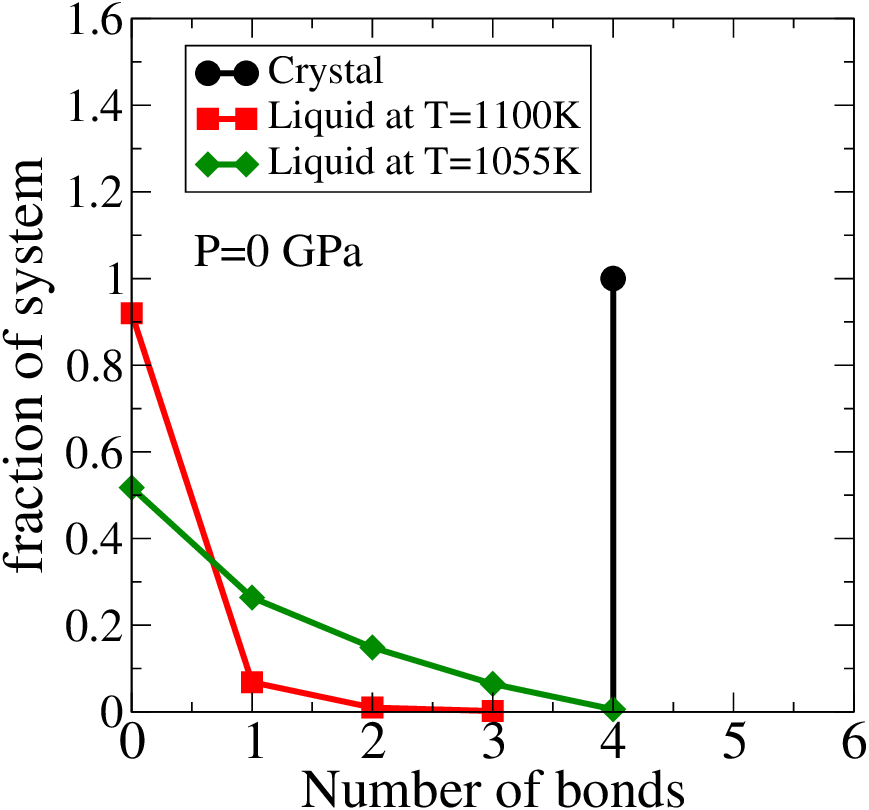}
    \caption{Typical distribution of number of connections per atom in high density liquid, low density liquid and in the pure crystalline phase.}
    \label{fig:fig-a3}
\end{figure}

The thermodynamics of whether the liquid is metastable with respect to crystallisation at a given temperature and pressure is determined by constructing the Landau free energy as a function of an order parameter. The order parameter(s) is chosen such that it distinguishes the liquid from the crystalline state sufficiently well. Here the strategy that is used is to identify crystalline particles and calculate the free energy cost to the growth of crystalline clusters of different sizes. Such an approach is broadly in consonance with Classical Nucleation Theory~\cite{debenedetti1996metastable,kashchiev2000nucleation} where the transition from the metastable liquid to the crystalline state occurs through rare fluctuations that generate crystalline clusters of different sizes. These clusters have a lower free energy in the bulk than the surrounding liquid, whereas the formation of an interface between the liquid and the solid induces a free energy cost. The bond orientational order parameters of Steinhardt and Nelson,\cite{steinhardt1983bond} $Q_{l}$ serve to distinguish local crystalline structures from disordered liquid ones. Specifically, the local analogue of these order parameters, $q_l$, can be used to distinguish the neighbourhoods of individual particles and classify them as being ordered or disordered. In terms of 
\begin{equation}
q_{lm}(i)=\frac{1}{n_b(i)}\sum_{j=1}^{n_b(i)}Y_{lm}[\theta(r_{ij}),\phi(r_{ij})],
\end{equation}
the order parameter $q_l$ is obtained by summing over $m's$:
\begin{equation}
q_l(i)=[\frac{4\pi}{(2l+1)}\sum_{m=-l}^l|q_{lm}(i)|^2]^{1/2}
\end{equation}
In the present work, we use $q_3(i)$ \cite{sastry2003liquid}, noting that using $q_6(i)$ is equally feasible, and gives very similar results~\cite{vasisht2013phase,vasisht2014nesting}.
The number of neighbours, $n_{b}(i)$, is taken to be the atoms within the first coordination shell of the pair-correlation function, i.e., atoms within a cut-off of $2.95~{\AA}$ from the reference atom.
Other works have considered other definitions, such as considering only the four nearest neighbours. However, when there are more than four atoms at similar distances from the reference atom, certain artefacts arise such as the apparent decrease of tetrahedral ordering with density or an increase with pressure~\cite{vasisht2014nesting}.
We therefore employ a distance-based cut-off to specify nearest neighbours. To identify crystalline particles, we compute the correlations in the local orientational order of neighbouring atoms, following the prescription described in the literature~\cite{van1992computer,ten1995numerical,wolde1996simulation,romano2011crystallisation,kesselring2013finite}. Atoms with correlated neighbourhoods of high local orientational order are classified solid-like particles.\\
Quantitatively, this correlation is given by,
\begin{equation}
Re \left( q_3(i).q_3(j)\right )=Re\left (\sum_{-3}^{3} q_{3m}(i)q_{3m}^*(j)\right )
\end{equation}
A particle $i$ and a particle $j$ are considered to be ``bonded" if $Re(q_3(i).q_3(j))<-0.23$ (see Fig.~\ref{fig:fig-a1}, Fig.~\ref{fig:fig-a2}). We note here the significance of the the cut-off value of $-0.23$ which demands that the crystal structure formed is diamond cubic, to the exclusion of the diamond hexagonal crystal structure which also has local tetrahedral ordering~\cite{romano2011crystallisation}.\\
Crystalline particles have a $q_3>0.6$ and are ``bonded" to at least 3 neighbours. Further, crystalline particles within the SW-cutoff distance of each other belong to the same cluster. In this study we employ  both the size of the largest cluster, $n_{max}$ and the full distribution of cluster sizes $P(n)$.\\
The distributions of $q_3$ and $Re(q_3(i).q_3(j))$ are shown in Fig.~\ref{fig:fig-a1} and Fig.~\ref{fig:fig-a2}. The distribution for the liquid at $T=1055K,~P=0~GPa$ is obtained from a non-crystallising MD trajectory of $90~ns$ and system size of $N=512$. The distribution for the liquid at $T=1100K,~P=0~GPa$ is obtained from a non-crystallising MD trajectory of $10~ns$ and system size of $N=512$. The histogram of number of bonded neighbours for the differently labelled particles is shown in Fig.~\ref{fig:fig-a3}. We note here that using $q_6(i).q_6(j)$ to identify crystalline particles gives nearly identical results when the appropriate cut-off is chosen\cite{vasisht2013phase}. The choice of cut-off will depend on whether a normalisation factor is included in the definition~\cite{ricci2019computational}.
\subsection{Kinetic reconstruction of free energy}\label{subsec:Appendix_MFPT}
The work here follows the method developed and described by the group of David Reguera where the kinetics of nucleation from a large number of free MD runs is utilised to obtain the free energy barrier to nucleation from the mean first passage time (MFPT)\cite{wedekind2007new,wedekind2008kinetic,wedekind2009crossover}.
The quantity that is often of most interest in the context of crystallisation is the crystallisation rate. For activated processes, which involve the crossing of a free energy barrier, the rate of crossing depends heavily on the height of the free energy barrier. In terms of a general reaction coordinate, $x$, this can be written as~\cite{kashchiev2000nucleation}
\begin{equation}
J_{cross}=\frac{1}{2}k ~e^{-\beta\Delta G(x^{*})}
\end{equation}
where $x^{*}$ is the value of the order parameter where the free energy $\beta \Delta G(x^{*})$ is maximum and $k$ is a kinetic pre-factor. In Classical Nucleation Theory, this free energy function is understood to have a dependence on the order parameter, $x$ in the following way:
\begin{equation}
\Delta G(x) = -\Delta \mu x + \sigma x^{2/3}    
\end{equation}
Here, $\Delta \mu$ is the difference in chemical potential between the bulk crystal and the bulk liquid and $\sigma$ is a term that describes the free energy cost due to the growth of the interface. Note that, when written in this form, the order parameter, $x$, is in fact the size of clusters and this equation describes the work of formation of different clusters of size $x$. Whereas the applicability of the form of $\Delta G(x)$ described above is debated, the work of formation is still broadly understood to include a net free energy gain in the bulk crystalline phase and a free energy cost to the growth of the interface between liquid and crystal, thus implying (under metastable conditions) that the free energy has a maximum.

In the description that follows, our goal is to calculate the free energy cost to the growth of clusters of size $x$, $\beta\Delta G(x)$, however, rather than to use the form described above, which requires the calculation of the chemical potential and the interfacial free energy, we use a method that relies on the kinetics of the process alone, without assumptions about the specific form of the free energy barrier.

The full theoretical basis for this method is excellently described in the original series of papers \cite{wedekind2007new,wedekind2008kinetic,wedekind2009crossover}. However, we summarize the formalism here for the sake of completeness. One can write the rate of crossing the nucleation barrier, assuming a diffusive crossing of an energy barrier which can be modelled using the Smoluchowski equation in one dimension, as follows. The process is described by the Smoluchowski equation
\begin{equation}
\begin{split}
{\partial P(x,t) \over \partial t} &= {\partial \over \partial x} \left[ D(x)e^{-\beta \Delta G(x)} \frac{\partial}{\partial x} (P_{st}(x)e^{\beta \Delta G(x)})\right] \\&= \frac{- \partial J(x,t)}{ \partial x}
\end{split}
\end{equation}
where $J(x,t)$ is the current, $D(x)$ is the order parameter dependent diffusivity,  $\Delta G(x)$ is the free energy function and $P(x,t)$ is the time dependent distribution function of the order parameter 
In the steady state, with $P(x,t) = P_{st}(x)$ we can write the expression for the rate of barrier crossing\cite{zwanzig2001nonequilibrium} as: 
\begin{equation}
J = -D(x)e^{-\beta \Delta G(x)} \frac{\partial}{\partial x} (P_{st}(x)e^{\beta \Delta G(x)}).
\end{equation}

In cases where the free energy $\Delta G(x)$ is not known, one can rearrange and arrive at
\begin{equation}
\beta \Delta G(x) = -ln P_{st}(x) - J \int \frac{dx'}{D(x')P_{st}(x')} + C
\label{eq:FP_bdg_int}
\end{equation}
One can also write the mean first passage time for a trajectory starting at $x_{0}$, to reach $x$, with a reflecting boundary condition $a$, and absorbing boundary at $b$\cite{zwanzig2001nonequilibrium}:
\begin{equation}
\tau(x;x_{0},a) = \int_{x_{0}}^{x} \frac{1}{D(y)}dy e^{\beta \Delta G(y)} \int_{a}^{y}dz e^{-\beta \Delta G(z)}
\label{eq:mfpt1}
\end{equation}
One can also write this in the following way:
\begin{equation}
\frac{\partial^2 \tau}{\partial x^2} = \left [\frac{\partial \beta \Delta G(x)}{\partial x} + \frac{\partial D(x)}{\partial x}\right]\frac{\partial \tau}{\partial x} + \frac{1}{D(x)} \nonumber
\end{equation}
This equation can be re-arranged and one can substitute $A(x) = \partial \tau (x)/\partial x$ to get, 
\begin{equation}
\frac{\partial ln(A(x)D(x))}{\partial x} = \frac{1}{D(x)A(x)} + \frac{\partial \beta \Delta G(x)}{\partial x}
\label{eq:AxDx}
\end{equation}
Further, one can write $B(x) = A(x)D(x)$ to get
\begin{equation}
\beta \Delta G(x) = ln(B(x)) - \int \frac{dx'}{B(x')} + C
\label{eq:1D_FE}
\end{equation}
From here, using $J=1/\tau(b)$ (where $\tau(b)$ is the mean first passage time at which the absorbing boundary at $b$ is reached), Eq.~\ref{eq:AxDx} can be combined with Eq.~\ref{eq:FP_bdg_int} to get
\begin{equation}
\frac{\partial (B(x)P_{st}(x))}{\partial x} = P_{st}(x) - \tau(b) \frac{\partial \tau(x)}{\partial x}
\end{equation}
Integrating this equation gives us,
\begin{equation}
\beta \Delta G(x) = \beta \Delta G(x=1) + ln \left ( \frac{B(x)}{B(1)}\right ) - \int_{1}^{x} \frac{dx'}{B(x')}
\label{eq:bdgvn_appendix}
\end{equation}
\begin{equation}
B(x) = -\frac{1}{P_{st}(x)}\left [ \int_{x}^{b} P_{st}(x')dx' - \frac{\tau(b)-\tau(x)}{\tau(b)}\right]
\label{eq:Bn_appendix}
\end{equation}
Eq.~\ref{eq:bdgvn_appendix} and Eq.~\ref{eq:Bn_appendix} are the equations used to reconstruct the free energy from the MFPT and the steady state probability. For the case of crystal nucleation, $x$ can be replaced with $n_{max}$, the size of the largest crystalline cluster, while $n_{max}=1$ is taken to be the second bin from the reflecting boundary condition. $\beta\Delta G(x=1)$ (or $\beta\Delta G(n_{max}=1)$) is an unknown constant at this point. Further, the small $n_{max}$ behaviour of involves special consideration.  We will discuss this issue and describe how the constant is determined in Appendix~\ref{subsec:nmax-v-n}.

The use of this method requires that the size of the largest cluster, $n_{max}$, be tracked in each of an ensemble of MD trajectories. From this, the steady state probability, $P_{st}(n_{max})$, and the mean first passage time, hereafter labelled $\tau_{MFPT}(n_{max})$, is calculated. Note that steady state here refers to the converged probability from an ensemble of MD trajectories rather than any steady state achieved in the trajectories. Each of the trajectories is extended till an absorbing boundary conditions is reached. An important practical aspect of using this method is that the ensemble of MD runs should preferably start from configurations with no crystalline ordering to effectively sample $\tau_{MFPT}(n_{max})$ and $P_{st}(n_{max})$ for the smallest possible $n_{max}$ values. This is especially important when one does not have the guarantee that unconstrained MD runs will fully sample the order parameter space starting from configurations with any arbitrarily chosen starting value of the order parameter. More details of the MD simulations performed for this method are given below in Appendix~\ref{subsec:MFPT_simulation_details}.
Our barrier calculations from MFPTs assume that the diffusion over the barrier satisfies the Smoluchowski equation, and hence that the trajectory over the barrier is continuous. Haji-Akbari has pointed out the need to carefully account for the fact that this assumption may not be satisfied for crystal nucleation, where the nucleus size can jump discontinuously due to attachment or detachment of clusters~\cite{haji2018forward}. The consistency between our kinetic and thermodynamic reconstruction of the free energy profiles for nucleation suggests that discontinuities due to attachment/detachment processes do not appear to matter for the systems and conditions we have studied in this work. Rationalising the underlying reasons would be  interesting to investigate further.  

This method has been used in a number of studies of nucleation\cite{lundrigan2009test,perez2011molecular,wedekind2015optimization,lu2015exploring}.

For a sufficiently high barrier, the mean first passage time, $\tau(x)$, is sigmoidal in form and can be used to extract information regarding the steady-state nucleation rate, the critical cluster size and the curvature at the top of the barrier, also known as the Zeldovich factor 
~\cite{wedekind2009crossover}. 
This aspect of the mean first passage time is not explored in this work; the barrier profile is used to determine the barrier height and critical cluster size. Moreover, in the general case, this method does not make any assumptions about the diffusivity of the order parameter, $D(x)$ or the shape of the energy barrier, $\beta\Delta G(x)$, beyond the overall framework of a diffusive barrier crossing in which the expression for $\tau(x)$ is written.
\subsubsection*{Simulation details}{\label{subsec:MFPT_simulation_details}}
The initial ensemble of configurations for the MD runs at each temperature of interest, at zero pressure, was prepared by first running a simulation in the isobaric, isothermal (NPT) ensemble at a high temperature of $T=1400K,~P=0~GPa$ and system size $N=512$ for $10~ns$. The MD runs are performed on the LAMMPS software suite using the velocity Verlet algorithm with a timestep of $0.3830$ fs\cite{plimpton1995fast}. Thermostatting and barostatting are done with a Nos$\acute{e} $-Hoover thermostat/barostat with time constants of $100$ and $1000$ steps respectively. These configurations have a mean density of $2.48g~cc^{-1}$ with standard deviation of $0.012g~cc^{-1}$. The relaxation time at these state points is of the order of $0.01~ns$ with diffusivities of the order of $10^{-4} cm^2/s$~~\cite{vasisht2013liquid}. After ignoring an initial transient, $600$ uncorrelated configurations were chosen as starting configurations. Energy minimisation was performed and the velocities were set to zero before being replaced with velocities corresponding to the target temperature. The length of the initial transient is chosen such that the liquid relaxes from the initial high temperature configuration. Subsequently, the liquid samples an initial metastable state corresponding to the target temperature, as discussed in Appendix~\ref{sec:Appendix_memoryLoss}. Each of the trajectories were simulated in the NPT ensemble using the velocity Verlet algorithm with the same timestep, thermostat and barostat. at the target temperature at $P=0~GPa$ till they crystallised. The first $0.04~ns$ were discarded and data gathered from the first time step after this where the total number of crystalline particles $n_{tot}=0$. This is to ensure that at $t=0$ the configurations are highly disordered with no crystalline ordering.  

An absorbing boundary condition is applied so that data is gathered only until the absorbing boundary is crossed for the first time. From this data, the free energy curve $\beta \Delta G(n_{max})$ is calculated using Eq.~\ref{eq:bdgvn_appendix} and Eq.~\ref{eq:Bn_appendix} with an Euler integration scheme 
Here, we emphasize that each of the independent trajectories needs to be extended till they reach the absorbing boundary. 
\subsection{Umbrella sampling}\label{subsec:Appendix_US}
The other technique used to determine the free energy cost to the growth of crystalline clusters is the umbrella sampling scheme which facilitates the reversible sampling of cluster sizes that are otherwise rarely sampled. 
The free energy cost to the growth of crystalline clusters of size $n$, $\beta G(n)$ is obtained (up to an additive constant) from the equilibrium probability density of sampling clusters of size $n$, with
\begin{equation}
    \beta\Delta G(n) = -ln(P(n)) + const.
    \label{eq:P_FE}
\end{equation}
Umbrella sampling is performed with NPT Monte Carlo simulations \cite{torrie1977nonphysical,frenkel2001understanding} to sample the desired range of order parameters  effectively. An in-house code was used for the umbrella sampling simulations that used an efficient double-sum implementation of the three-body Stillinger Weber potential described in Ref.~\onlinecite{saw2009structural} and Ref.~\onlinecite{makhov2003isotherms}. 
A standard Metropolis scheme is used for the Monte Carlo (MC) simulations with an MC step consisting of either $N$ single particle trial displacements or a trial change in the volume.
Trial displacements are accepted or rejected with a probability
\begin{equation}
P_{accept}(o \rightarrow n) = min\{1,exp\left[-\beta (E_{n}-E_{o})\right]\}
\end{equation}
Trial changes in the volume are accepted with the probability
\begin{widetext}
\begin{equation}
P_{accept}(o \rightarrow n) = min\{1,exp\left[-\beta \left[(E_{n}-E_{o}) +P(V_{n}-V_{o})\right] + (N+1)ln\frac{V_{n}}{V_{o}}\right]\}
\end{equation}
\end{widetext}
Two variants of this method are used, both of which involve the imposition of a bias potential on the size of the largest cluster, $n_{max}$, based on previous work \cite{ten1995numerical}. 
However, as will be discussed in forthcoming sections, in general, $P(n_{max})~\neq~P(n)$ except under certain conditions. This issue has been discussed in the literature as well~\cite{ten1997enhancement,saika2006test,maibaum2008comment,wedekind2009crossover,lundrigan2009test}. The general expression for the Hamiltonian under application of bias is given by:
\begin{equation}
H_C = H + W(n_{max})
\label{eq:bias_HW}
\end{equation}
where $W(n_{max})$ represents the bias potential and $H$ is the original Hamiltonian.\\
In the first instance, a harmonic bias of the form 
\begin{equation}
W(n_{max}; n_{max}^{0},k_{n_{max}}) = {1 \over 2} k_{n_{max}} (n_{max} - n_{max}^{0})^2
\label{eq:W_harmonic}
\end{equation}
is used to enhance sampling around a desired value of $n_{max}$, labelled $n_{max}^{0}$.
The sampling of different values of $n_{max}$ is enhanced by running multiple simulations with each independent simulation having a different bias centre, $n_{max}^{0}$, or bias potential $W(n_{max}; n_{max}^{0})$, thus sampling different windows of $n_{max}$ values.
\par
In order to address the complications 
arising from the choice of a harmonic bias on the order parameter~\cite{saika2006test}, $n_{max}$, we also consider a different bias protocol for the umbrella sampling scheme when attempting to measure the free energy barrier at deep supercooling. Here, the Hamiltonian is modified by adding a constraining potential of the hard wall form rather than a harmonic bias, as described by Saika-Voivod et al\cite{saika2006test}. 
The hard wall bias strictly constrains the size of the largest cluster to be between $n_{max}^l$ and $n_{max}^u$ as described in Eq.~\ref{eq:W_HW}. Different independent simulations constrain sampling within different bounds. The full cluster size distribution is also used, from which we can calculate free energy using Eq.~\ref{eq:P_FE}. For the purposes of comparison, the free energy as a function of $n_{max}$ is also calculated from simulations with the hard wall bias.\\
The corresponding bias potential then takes the form:
\begin{align}
W &=
\begin{cases}
0 & n_{max}^l \leq n_{max} < n_{max}^u\\
\infty & otherwise
\end{cases}
\label{eq:W_HW}
\end{align}
To improve equilibration, we perform parallel tempering, wherein simulations at different temperatures or for different bias potentials ($n_{max}^{0}$  or $[n_{max}^l,n_{max}^u]$ values) are run in parallel, and configurations for distinct temperatures or bias potentials (with adjacent values of $n_{max}^{0}$  or $[n_{max}^l,n_{max}^u]$)  are swapped periodically, using the parallel tempering Metropolis scheme.\\
Short segments of the trajectory of $50$ MC steps are generated with the unbiased Hamiltonian. These are then accepted or rejected with a probability,
\begin{equation}
P_{accept}(o \rightarrow n) = min\{1,exp\left[-\beta  (W_{n} - W_{o})\right]\}
\end{equation}
In this case, the $o$ and $n$ configurations refer to those at the beginning of the trajectory segment and at the end, respectively. Note that for the case of the hard wall bias, $W_{n}-W_{o}$ is either $0$ or $\infty$.
These simulations are used to generate a distribution of $n_{max}$ values $P_b(n_{max})$, where the subscript $b$ refers to sampling in the biased ensemble. One can obtain the unbiased distribution of $n_{max}$ (up to normalisation) using the relation
\begin{equation}
	P(n_{max})= P_b(n_{max}) e^{\beta W(n_{max})}
	\label{eq:P_nmax}
\end{equation}
where $P(n_{max})$ is the frequency with which the largest cluster samples a size, $n_{max}$.\\
From the unbiased distribution, one obtains the Landau free energy as a function of $n_{max}$ as: 
\begin{equation}
\beta\Delta G(n_{max}) = -ln(P(n_{max})) + const.
\label{eq:nmax_FE}
\end{equation}
From Eq.~\ref{eq:nmax_FE}, we wish to identify the constant the yields $\beta\Delta G(0)=0$.
The estimates are obtained from simulations with different $n_{max}^0$ or $[n_{max}^l,n_{max}^u]$ bounds and sample different but overlapping windows of $n_{max}$.
Here, we make a distinction between the free energy calculated at high temperatures, where the critical cluster is expected to be large, and free energies calculated when the critical cluster is expected to be small. In the former case, the missing constant in each independent simulation, specified by index $d$, is obtained by fitting $\beta\Delta G_d(n_{max})$ to a single polynomial of $n_{max}$. This is done by a least square fit, by minimising 
\begin{equation}
\begin{split}
	\chi_{US} = \sum_{d=1}^{N_{sim}} \sum_{n_{max}=n_{lo}^d}^{n_{hi}^d} \Biggl [ \beta\Delta G_d(n_{max}) + a_0n_{max} \Biggr. \\  \Biggl . -a_1n_{max}^{2/3}  - \sum_{i=2}^{p} (a_i n_{max}^i) - b_d  \Biggr ]^2
	\end{split}
\end{equation}
where $N_{sim}$ is the number of independent simulations and $n_{lo}^d$ and $n_{hi}^d$ are respectively the lower and upper bounds within which $n_{max}$ is sampled in the simulation indexed $d$. The index $d$ runs from $1$ to $N_{sim}$. 
$p$ is the order of the polynomial with coefficients $a_i$, and $b_d$ will give us the missing constants.
The CNT expression $\beta\Delta G(n) = -a_0 n + a_1 n^{2/3}$ can be expected to be valid for sufficiently large critical clusters and high free energy barriers. Hence a polynomial of the form $a_0 n + a_1 n^{2/3} + a_2 n^2 + a_3 n^3 \cdots$ is used.

Where the critical cluster is expected to be small, we make no assumption of a CNT-like polynomial fit. The overlap between $\beta\Delta G(n_{max})$ obtained from different simulations sampling adjacent bounds is maximised by identifying the appropriate constant for each independent simulation, $b_d$. This is done by minimising the following quantity:
\begin{equation}
\begin{split}
\chi_{HW} = \sum_{d=1}^{N_{sim}} \sum_{n_{max}=n_{lo}^d}^{n_{hi}^d} \Bigg[ \beta\Delta G_d(n_{max}) \Biggr. \\ \Biggl. - \beta\Delta G_{d+1}(n_{max}) 
- b_d \Biggr ]^2 \label{eq:HW_stitch_nmax}
\end{split}
\end{equation}
$N_{sim}$ is the number of independent simulations and $[n_{lo}^d,n_{hi}]^d]$ is the range of $n$ over which adjacent simulations overlap. The index, $d$, runs over each independent simulation, starting from $d=1$ to $d=N_{sim}$. This procedure yields the free energy, $\beta\Delta G(n_{max})$, up to an unknown constant as given in Eq.~\ref{eq:nmax_FE}.
As mentioned for the reconstruction of $\beta\Delta G(n_{max})$ through the MFPT approach, as well as umbrella sampling, the procedure used to determine this remaining unknown constant is described in Appendix~\ref{subsec:nmax-v-n}.
\par
The umbrella sampling runs using the hard wall bias are also used to obtain $\beta\Delta G(n)$. $\beta\Delta G(n)$ can also be obtained from $N(n)$ from umbrella sampling runs with a harmonic bias, which we do not do here. The unbiased expectation value of $N(n)$ (the number of clusters of size $n$) can be written as:
\begin{equation}
\left < N(n) \right >  = \frac{\left < N(n) e^{\beta W}\right >_C }{\left < e^{\beta W)}\right >_C}
\label{eq:unbias_HW}
\end{equation}
The expectation subscript $C$ is the sampled probability from the simulation under the modified Hamiltonian.
The un-biasing described in Eq.~\ref{eq:unbias_HW} simplifies since $W=0$ or $W=\infty$ depending on the size of the largest cluster. For the case of the hard wall bias potential, one can thus replace $\left < N(n) \right > = \left < N(n) \right >_C$ within the constrained region.

We compute $\beta\Delta G(n)$ from $-ln\left(N(n)\right)$ up to an unknown constant, within the window in which we perform biased sampling.
We use the equilibrium data of $P(n) = N(n)/N(0)$ at small $n$  and demand that $P(n)$ from simulations sampling other values of $n$ sequentially match these, as described by Eq.~\ref{eq:HW_stitch_n}. 
From a set of independent simulations, each indexed by $d$ and having distinct but adjacent bounds, one obtains the free energy differences $\beta\Delta G_d(n)$ up to an undetermined constant, $b_d$. The constants, $b_d$, are obtained by minimising the error described in Eq.~\ref{eq:HW_stitch_n}, $\chi_{HW}$, sequentially between overlapping data points from simulations with adjacent bounds.  
\begin{equation}
\chi_{HW} = \sum_{d=1}^{N_{sim}} \sum_{n=n_{lo}^d}^{n_{hi}^d} \left [\beta\Delta G_d(n) - \beta\Delta G_{d+1}(n) - b_d \right ]^2
\label{eq:HW_stitch_n}
\end{equation}
This is done in the same way as in Eq.~\ref{eq:HW_stitch_nmax}, but subject to the constraint $\quad \beta\Delta G_d(0)=0~\quad$ if $n_{lo}^{d}=0$.
Note that unlike the procedure in Eq.~\ref{eq:HW_stitch_nmax} for $\beta\Delta G(n_{max})$, the added constraint in Eq.~\ref{eq:HW_stitch_n} that $\beta\Delta G_d(n=0)=0$ if $n_{lo}^{d}=0$ does not leave behind an undetermined additive constant.
Another important point here is that no assumption to fit the CNT form is made; $\beta\Delta G(0)=0$ as a consequence of how quantities have been defined.
\subsubsection*{Simulation details}
Umbrella sampling Monte Carlo simulations in the NPT ensemble were started by first randomly placing $N$ particles in a box, taking care to prevent any two particles being too close so that large repulsive interactions are avoided. The initial box size corresponded to a density of $2.48g~cc^{-1}$. For simulations at deep supercooling, where the hard wall bias is applied, simulations were initially equilibrated with a harmonic bias potential for $10^7$ MC steps with a spring constant of $k_{n_{max}}=0.01\epsilon$. The harmonic bias was replaced with the hard wall bias after the initial equilibration under a harmonic constraint, taking care that the cluster size in each window be within the desired bounds. Thereafter, $5\times10^6$ MC steps were performed with the hard wall bias before statistics were gathered for a subsequent $2.5\times10^7$ MC steps.
The auto-correlation functions of density ($\rho$), $Q_6$ and potential energy were monitored under the application of the hard wall bias and the relaxation time found to be similar and less than $10^5$ MC steps for all the windows and for each of the three quantities considered. Thus, keeping in mind a relaxation time of $\tau=10^5$ MC steps, we use an equilibration length of $50\tau=$ and a production length of $250\tau$.

Hard wall constraints are placed at $(n_{max}^l,n_{max}^u) = [0,2],[1,3],[2,4] \dots$ with parallel tempering swaps performed between simulations with adjacent and overlapping constraint, or adjacent temperatures, to speed up equilibration (see Appendix~\ref{subsec:Appendix_PT}).
At state points where the free energy barrier is expected to be high, and the liquid is unambiguously metastable, a number of independent NPT MC simulations are initialised, each constraining $n_{max}$ in the vicinity of some $n_{max}^{0}$ with the use of a harmonic potential with spring constant $k_{n_{max}}$. Each independent simulation
is equilibrated for $10^6$ MC steps or $10\tau$. Parallel tempering swaps were performed between simulations with adjacent $n_{max}^0$, or adjacent temperatures, to speed up equilibration. 
The length of the production run over which the order parameters are sampled is determined in the following way. Each simulation is assigned a bias potential, specified either by the bias center or the bounds ($n_{max}^0$ or $[n_{max}^l,n_{max}^u]$), as well as a temperature. Parallel tempering exchanges result in swaps between simulations with adjacent temperatures or bias potentials.
This process should result in each simulation, with a ``native" temperature and bias potential, ``visiting" every other temperature or bias potential a finite number of times. We measure the time taken for the simulation with the lowest temperature or bias potential to visit the highest temperature or bias potential $10$ times. The length of the production run is taken as the number of MC steps required for $10$ such exchanges to happen, along each of the two axes, temperature and bias potential. An exception to this is when the number of MC steps taken for $10$ exchanges to occur is less than $10^7$ MC steps, in which case the production length is taken to be $10^7$ MC steps.

At these state points, umbrella sampling with a harmonic bias is used, taking statistics on $n_{max}$, to construct the free energy curves. Additional runs with a hard wall bias on $n_{max}$ are performed where statistics for $P(n)$ are obtained for the smallest cluster sizes. This is done to enhance sampling near $n=0$, and to avoid the issues described below in Sec.~\ref{subsec:nmax-v-n}.
\subsection{Parallel Tempering}\label{subsec:Appendix_PT}
The general expression for probability of acceptance of parallel tempering swaps in the NPT ensemble between simulations indexed $i$ and $j$ is given by
\begin{equation}
\begin{split}
P_{accept} = min\Biggl(1, exp\Bigl[\bigl[ (E_i -E_j) + P(V_i - V_j)\bigr] (\beta_i - \beta_j)\Bigr]\Biggr. \\
\Biggl. exp\bigl[-\beta_j W_i(n_{max_j}) - \beta_i W_j(n_{max_i})\bigr]\Biggr.\\
\Biggl. exp\bigl[\beta_i W_i(n_{max_i}) + \beta_j W_j(n_{max_j}) \bigr]\Biggr )
\end{split}
\end{equation}
The details of parallel tempering are as follows:
\begin{itemize}
\item Consider $N$ independent simulations run in parallel - different temperatures or different bias potentials.
\item To ensure better sampling of the phase space {\bf r(t)} and consequently of the order parameter, we swap adjacent configurations periodically.
\item Two types of swaps are performed, one type where simulations with different temperatures but the same bias potential exchange configurations and one type where simulations at the same temperature but different bias potentials exchange configurations.
\item A swap between adjacent simulations indexed $i$ and $j$, at different temperatures, $1/\beta_i$ and $1/\beta_j$, but with the same bias potential is executed with a probability of $min\Biggl ( 1,exp\bigl[ (E_i -E_j) + P(V_i - V_j)\bigr] (\beta_i - \beta_j) \Biggr )$
\item For cases where $\beta$ is the same but the bias potential varies, the probability is $min\left ( 1,exp\left[ \beta(W_N - W_O) \right] \right )$
\item Here, the term $W_N - W_O$ represents the sum of the bias potentials after the swap minus the sum of the bias potentials before the swap (the sum being over the bias applied on the two runs in consideration.
\begin{eqnarray}
W_N &=& W_j(n_{max_i}) + W_i(n_{max_j}) \nonumber \\
W_O &=& W_i(n_{max_i}) + W_j(n_{max_j}) \nonumber
\end{eqnarray}
For the hard wall bias, the swap is accepted with probability $1$ if the $n_{max_i}$ and $n_{max_j}$ are both within the new constraints after the swap and rejected otherwise.
\end{itemize}
\subsection{Consistency of free energy reconstructions at small cluster sizes }\label{subsec:nmax-v-n}
In computing the free energy barrier to nucleation, the size of crystalline clusters, $n$, is employed as the order parameter, and the equilibrium probability density of cluster sizes, $P(n)$ is related to the free energy cost to the formation of a crystalline cluster of size $n$, by Eq.~\ref{eq:P_FE}. Using Eq.~\ref{eq:HW_stitch_n} subject to the constraint that $\beta\Delta G(0)=0$ allows us to relate $\beta\Delta G(n)$ to $-ln\left(P(n)\right)$ without any unknown constants.

Often, (including parts of the present work) the order parameter, $n_{max}$, and the corresponding distribution, $P(n_{max})$, is used as a proxy to $P(n)$. The use of $n_{max}$ as the order parameter describing the crystallisation transition is appropriate only when $P(n_{max})=P(n)$. \cite{wolde1996simulation,saika2006test,maibaum2008comment,wedekind2009crossover,lundrigan2009test}. 
In a finite volume, the statistics of the largest cluster, $n_{max}$, often show that configurations containing a small cluster (i.e., where the largest cluster is small) are more frequently sampled than configurations where there are no crystalline particles at all.
As mentioned earlier in Section~\ref{sec:methods}, this leads to the appearance of an artificial minimum in $\beta\Delta G(n_{max})$ at small values of $n_{max}$ as discussed at length in \cite{saika2006test,chakrabarty2008chakrabarty,maibaum2008comment,wedekind2009crossover,lundrigan2009test}. This effect is more pronounced at deeper supercooling and larger system sizes as shown (and later discussed) in Fig.~\ref{fig:fig-1} where the deviation between the largest cluster distribution, $P(n_{max})$, and the full cluster size distribution, $P(n)$, is significant~\cite{maibaum2008comment,wedekind2009crossover}.

For clusters larger than a size $n_{low}$, such that clusters of size $n_{low}$ are rare, $P(n_{max})=P(n)~\forall~n,n_{max}\geq n_{low}$ \cite{saika2006test,auer2004numerical,reiss1999some}. Here, rare clusters are those for which the frequency with which clusters of size $n_{low}$ are observed is well-approximated by the probability of observing one such cluster, and for which the formation of multiple such clusters can be considered independent events.
In this limit, $P(n_{max})$ does not display system-size dependence, while for smaller clusters ($n_{max}<n_{low}$), system-size dependence is apparent.
A different system size effect is evident when considering state points with large critical clusters whose diameter is greater than half the box length, thus inducing ordering across periodic images (see Appendix~\ref{subsec:highT_sysSize}).\\

On the other hand, $P(n)$ is independent of system size for all $n$. Given $\beta\Delta G(n_{max})$ up to an unknown additive constant, the question is how this relates to $\beta\Delta G(n)$. One uses the fact that $\beta\Delta G(n)=-ln(P(n))$.
However, $P(n_{max})=e^{-\beta\Delta G(n_{max})}$ is known to deviate from $P(n)$ for small $n$, up to some (as yet unknown) cluster size $n_{low}$.
For $\beta\Delta G(n_{max})$ obtained from umbrella sampling runs, we employ the procedure of using the equilibrium distribution $P(n)$ to define our estimate of $\beta\Delta G(n_{max})$ up to an $n$ value $n_{hi}>n_{low}$, and demanding that $\beta\Delta G(n_{max} = 0) = 0$. 
In the case of the MFPT runs, we make the reasonable assumption that the steady state probability of observing clusters of size $n$, $P_{ss}(n)=\left<N(n)/N(0)\right>$, is equal to the equilibrium probability, $P(n)$, for $n~\leq~n_{hi}$. Here, $n_{hi}~\geq~n_{low}$ is an as yet unknown upper limit up to which this assumption holds and the average is over an ensemble of independent, unconstrained MD trajectories. Note that $P_{ss}(n)$ is not the steady state probability of sampling the largest crystalline cluster, $P_{st}(n_{max})$. An explicit comparison is made in Fig.~\ref{fig:fig-1} to show that this approximation holds for some $n_{hi}$.
On the other hand, for the umbrella sampling runs, we obtain $P(n)$ from $N(n)/N(0)$ as described before.
This procedure is represented by the expression in Eq.~\ref{eq:HW_mfpt_compare} where we require that $\beta\Delta G(n_{max})=-ln(P(n))~\forall~n_{max}~\leq~n_{low}$ and make the demand that $\beta\Delta G(n_{max})~\approx~-ln(P(n))$ for $n_{low}~\leq~n_{max}~\leq~n_{hi}$. The following error is then minimised:
\begin{equation}
\chi_{c}=\sum_{n,n_{max}=n_{low}}^{n_{hi}}\left|\beta\Delta G(n_{max})+ln(P(n)) + C\right| \delta_{n,n_{max}}
\label{eq:HW_mfpt_compare}
\end{equation}
Here, the sums are over $n$ and $n_{max}$, considering only those terms where $n=n_{max}$. The unknown constant $C$ that minimizes the difference between $\beta\Delta G(n_{max})$ and $-ln(P(n))$ within the range $[n_{low},n_{hi}]$is determined. The choice of $n_{low}$ and $n_{hi}$, as motivated by the discussion above, is determined by the deviation of $P(n_{max})$ from $P(n)$ at small $n$ or $n_{max}$, as well as the limit up to which the equilibrium $P(n)$ is well-approximated by $P_{ss}(n)$.

Similar methods have been used in Ref.~\onlinecite{wedekind2009crossover} and Ref. ~\onlinecite{lundrigan2009test}.
The applicability of the procedure described in this section has limits if $n_{low}$ itself shifts to values comparable to the critical size $n^{*}$. At lower temperatures, as $P(n_{max})$ and $P(n)$ become progressively more different and the appropriate $n_{low}$, beyond which $P(n_{max}~\approx~P(n)$, shifts to larger values, this comparison between $\beta\Delta G(n)$ and $\beta\Delta G(n_{max})$ becomes more difficult to the point that it is eventually no longer tenable.
\renewcommand{\thefigure}{B\arabic{figure}}
 \renewcommand{\theequation}{B\arabic{equation}}
  \setcounter{figure}{0}
\section{Free energy reconstruction with $Q_6$ as the order parameter}\label{section:Appendix_Q6_MFPT_US}
We perform free energy reconstructions to obtain $\beta\Delta G(Q_6)$ using both the mean first passage time method as well as umbrella sampling.
We define the global bond orientational order, $Q_6$ in the following way. We first define the global parameters $Q_{lm}$ as
\begin{equation}
Q_{lm}=\frac{1}{N_b}\sum_{i=1}^N\sum_{j=1}^{n_b(i)}Y_{lm}[\theta(r_{ij}),\phi(r_{ij})].
\end{equation}
Here, $n_b(i)$ is the number of neighbours for the $i^{th}$ particle. $Y_{lm}$ is the spherical harmonic and $N_b$ is the total number of neighbor pairs. neighbours are defined to be particles separated by a distance less than the first minimum of the radial distribution function. The global bond orientational order can now be expressed in terms of $Q_{lm}$
\begin{equation}
Q_l=[\frac{4\pi}{(2l+1)}\sum_{m=-l}^l|Q_{lm}|^2]^{1/2}
\end{equation}
For the mean first passage time method, we use the same set of $600$ NPT MD crystallising trajectories of $N=512$ particles to sample $P_{st}(Q_6)$ and $\tau_{MFPT}(Q_6)$.
One can identify the liquid basin at $Q_6=0.03$ from $P_{st}(Q_6)$. Data from each trajectory is gathered from when $Q_6$ first attains a value of $0.03$ until an absorbing boundary condition at $Q_6=0.12$ is reached. The absorbing boundary is chosen to be clearly on the crystalline side of the free energy barrier~\cite{ricci2019computational}.

Umbrella sampling MC simulations are performed in the NPT ensemble with $N=512$ particles at $T=1070K$, $P=0~GPa$ with a harmonic bias on $Q_6$. Parallel tempering swaps between adjacent bias windows are performed.
The auto-correlations of density and potential energy are found to decay over a timescale of $5\times10^5-10^6$ MC steps. Equilibration is performed for $15\times10^6$ MC steps with a subsequent production run of $15\times10^6$ MC steps.

\begin{figure}[htpb!]
    \centering
    \includegraphics[scale=0.5]{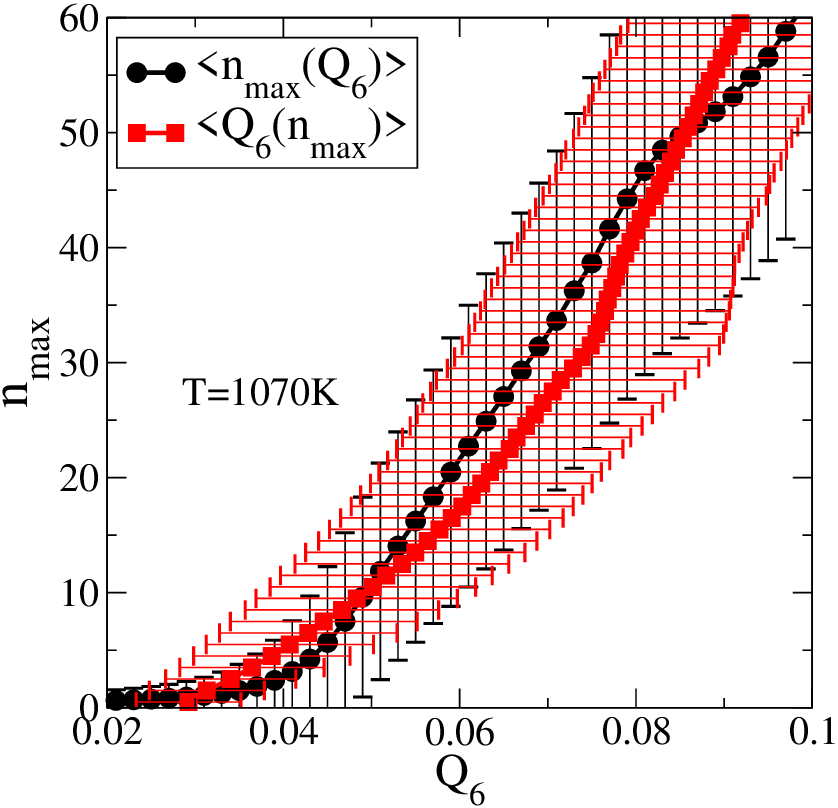}
    \caption{The parametric dependence of $n_{max}$ on $Q_6$ and vice versa. Data gathered from the set of $600$ independent unconstrained NPT MD runs of $N=512$ particles at $T=1070K$, $P=0~GPa$. Error bars indicate standard deviations. We use $\langle n_{max}(Q_6)\rangle$ to compute the derivative in Eq.~\ref{eq:bDG_map} and map the free energies.}
    \label{fig:fig-b1}
\end{figure}

\begin{figure}[htpb!]
\centering
\includegraphics[scale=0.5]{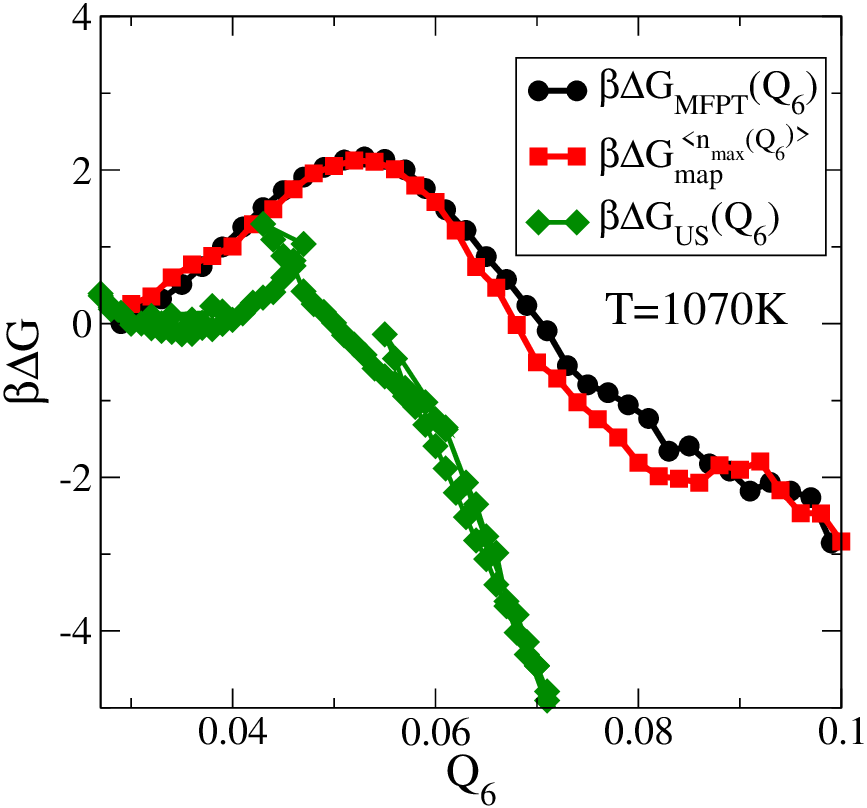}
\caption{The free energy reconstruction as a function of $Q_6$ using the MFPT reconstruction, labelled, $\beta\Delta G_{MFPT}(Q_6)$ is compared with the reconstruction using umbrella sampling, labelled $\beta\Delta G_{US}(Q_6)$. Additionally, the parametric dependence of $n_{max}$ on $Q_6$, computed from the average $\langle n_{max}(Q_6)\rangle$ is used to map order parameters and use $\beta\Delta G(n_{max})$ (see Fig.~\ref{fig:fig-1}) to obtain $\beta\Delta G_{map}^{\langle n_{max}(Q_6)\rangle}(Q_6)$. }
\label{fig:fig-b2}
\end{figure}

One can check for consistency between $\beta\Delta G(n_{max})$ and $\beta\Delta G(Q_6)$ by expressing $n_{max}$ as a parametric function of $Q_6$. From each point on a trajectory, one obtains an $n_{max})$ and a $Q_6$. One can compute the average $n_{max}$ corresponding to a given $Q_6$, $\langle n_{max}(Q_6)\rangle$ by aggregating data over all points on a trajectory and over multiple trajectories. One can also do the converse to get $\langle Q_6(n_{max}) \rangle$. In Fig.~\ref{fig:fig-b1}, the parametric dependence is shown from the set of crystallising NPT MD trajectories.
One can relate $\beta\Delta G(n_{max})$ to $\beta\Delta G(Q_6)$ in the following way by considering that the equilibrium distributions can be related by
\begin{equation}
    P_{eq}(Q_6)dQ_6 = P_{eq}(n_{max})dn_{max}
\end{equation}
From this, we can write
\begin{equation}
    \beta\Delta G_{map}(Q_6) = \beta\Delta G(n_{max}) - ln\left| \frac{dn_{max}}{dQ_6} \right|
    \label{eq:bDG_map}
\end{equation}
The comparison between $\beta\Delta G(Q_6)$ and $\beta\Delta G_{map}(Q_6)$ is made in Fig.~\ref{fig:fig-b2} for $T=1070K$, $P=0~GPa$ to demonstrate the point.
One observes that while the MFPT reconstruction shows a barrier at $T=1070K$ that can be shown to be consistent with $\beta\Delta G(n_{max})$, $\beta\Delta G_{US}(Q_6)$ does not. Moreover, $\beta\Delta G_{US}(Q_6)$ will clearly not show similar consistency under a transformation of variables. We stress that $\beta\Delta G(n_{max})$ gives results that are in quantitative agreement at all the state points considered regardless of the choice of order parameter. Another point worth mentioning is that estimates for $\beta\Delta G_{US}(Q_6)$ from overlapping regions for adjacent bias windows do not match under the conditions described here suggesting that the configurations that are sampled do not have a one-to-one correspondence with the value of $Q_6$. 
We note that the issues described herein are of particular relevance when using $Q_6$ to estimate the barrier between the liquid and the crystalline state. 
However, the free energy difference between the liquid and the fully crystalline state is accurately estimated using $Q_6$, as discussed in the context of water in Ref.~\onlinecite{palmer2014metastable}, which contains internal consistency checks for the free energy difference between the metastable liquid and crystalline basins. Further, the liquid-liquid coexistence conditions in Ref.~\onlinecite{palmer2014metastable}, from free energy calculations, are consistent with 2-state model calculations in~\cite{holten2014two}.
\renewcommand{\thefigure}{C\arabic{figure}}
 \renewcommand{\theequation}{C\arabic{equation}}
  \setcounter{figure}{0}
\section{Variation of the properties of the metastable liquid with temperature}\label{sec:Appendix_memoryLoss}
\begin{figure}[htpb!]
    \centering
    \includegraphics[scale=0.42]{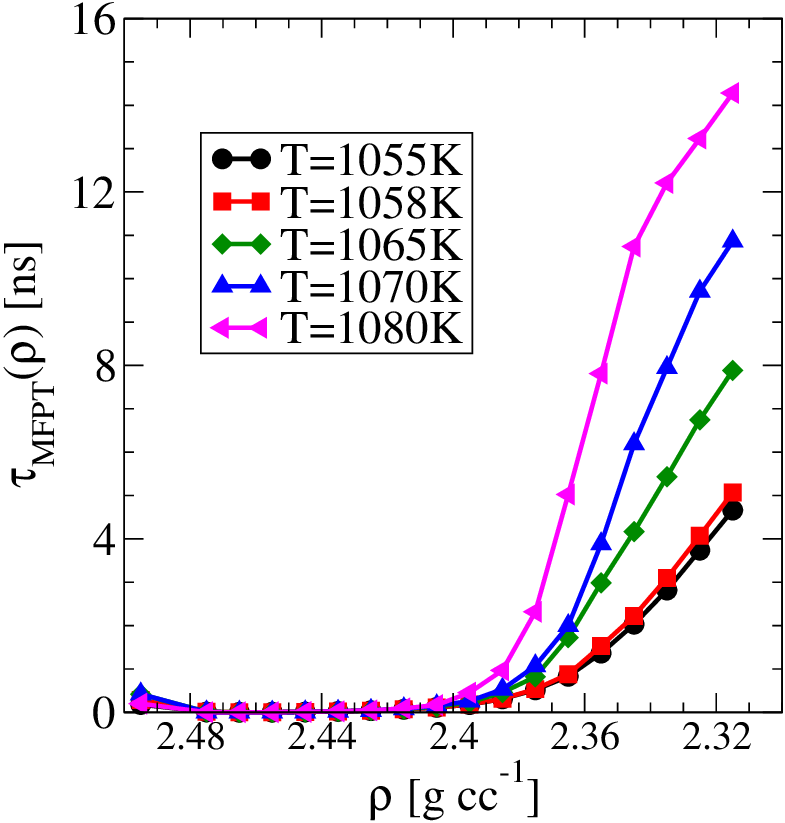}
    \caption{The mean first passage time of the density, $\rho$ from a set of independent MD runs at five temperatures. At each temperature, $600$ independent NPT MD runs were conducted at $P=0~GPa$ and a system size of $N=512$.}
    \label{fig:fig-c1}
\end{figure}

The mean first passage time of the density, $\rho$, shows that the liquid explores high density configurations on the timescale of $<1~ns$ before the density begins to drop (see Fig.~\ref{fig:fig-c1}). To understand the initial metastable state into which the liquid settles, we observe the change in the density profile of the metastable liquid with temperature during the first $0.5~ns$. We ignore the first $0.038~ns$ as a transient during which the thermostat and barostat come into effect. 
The dependence of the mean density on the target temperature is compared, and the procedure is repeated for inherent structure energies as well. Given that the statistics are similar with a similar trend, regardless of the ensemble of starting configurations, this suggests that the sampled configurations are independent of starting configurations. An important point is that both sets of initial configurations are highly disordered with no crystalline ordering -- one may not expect that the liquid will relax to the same initial metastable state if the initial configuration already has significant crystalline ordering.
\begin{figure}[htpb!]
    \centering
    \includegraphics[scale=0.42]{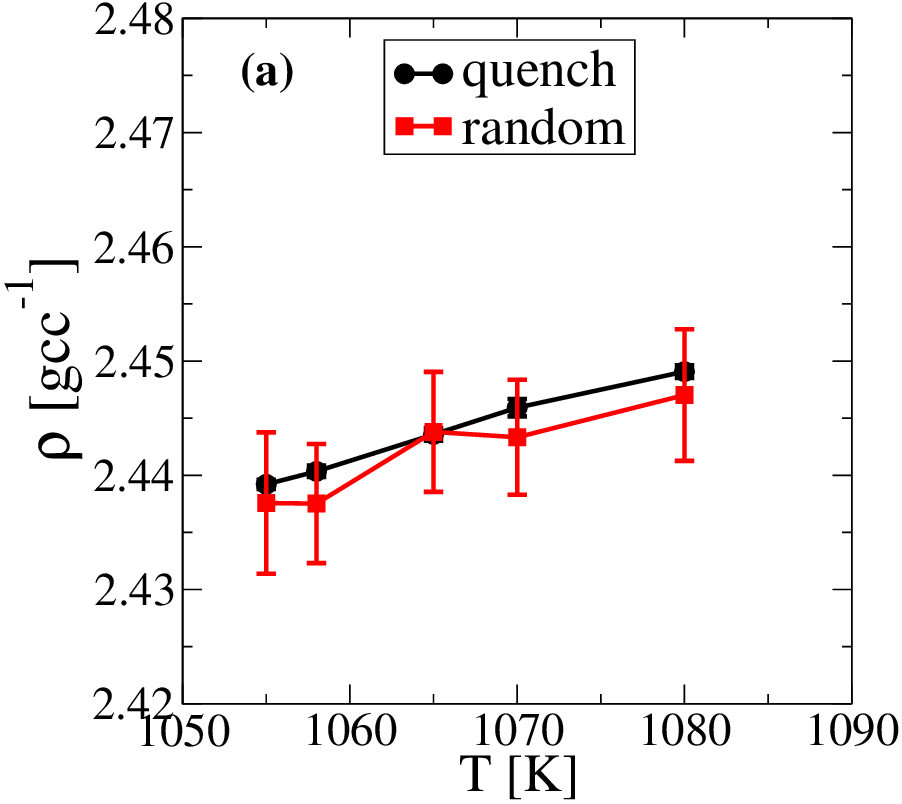}
    \includegraphics[scale=0.42]{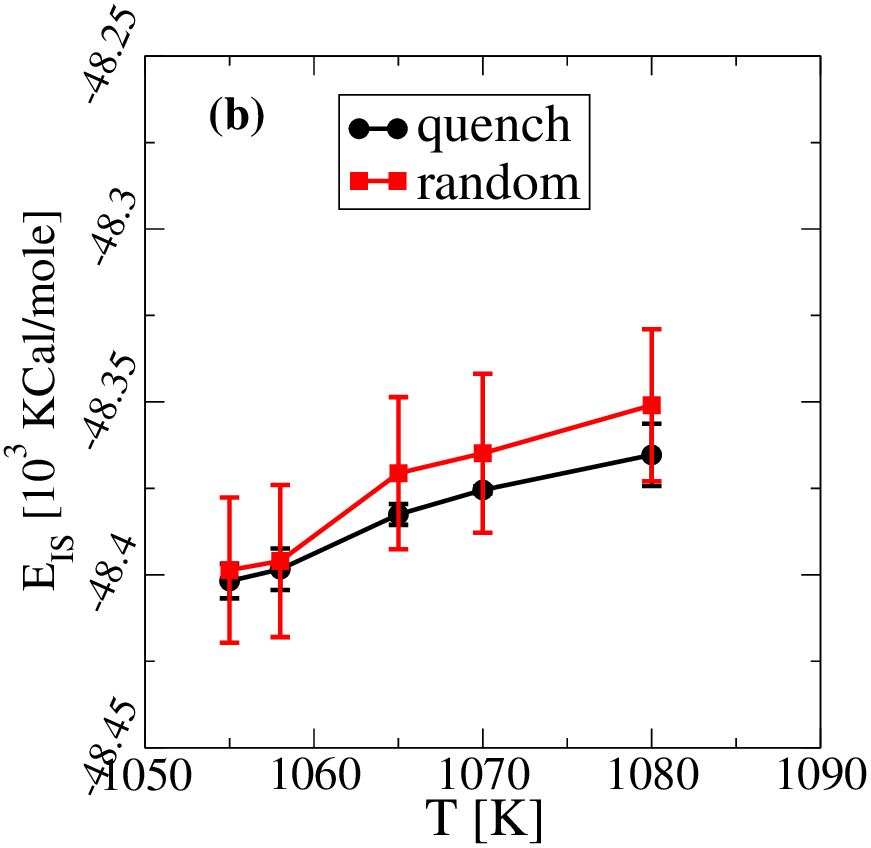}
    \caption{Average density ({\it (a)}) and inherent structure energy ({\it (b)}) vs temperature at $P=0~GPa$ and $N=512$ from two sets of starting conditions -- $8$ from random initialisation and $600$ from the quench. The initial $0.038~ns$  were discarded as the initial transient and the next $0.5~ns$ were used. The error bars represent the uncertainty in the mean.}
    \label{fig:fig-c2}
\end{figure}
\renewcommand{\thefigure}{D\arabic{figure}}
 \renewcommand{\theequation}{D\arabic{equation}}
  \setcounter{figure}{0}
\section{System size effects for large critical cluster}\label{subsec:highT_sysSize}
\begin{figure}[!htpb]
\centering
\includegraphics[scale=0.4]{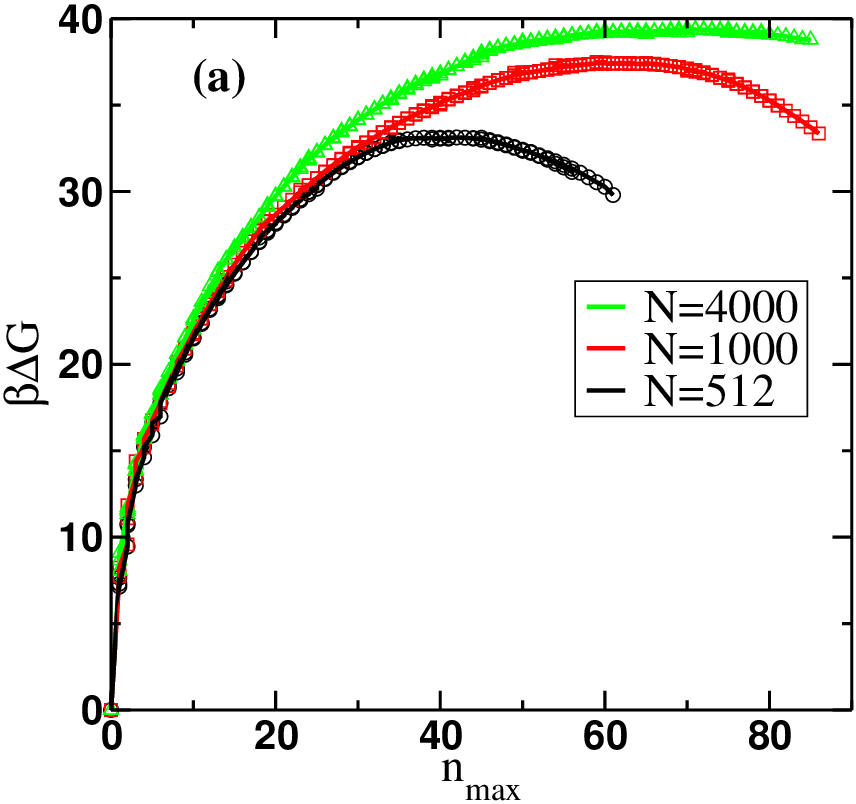}
\includegraphics[scale=0.4]{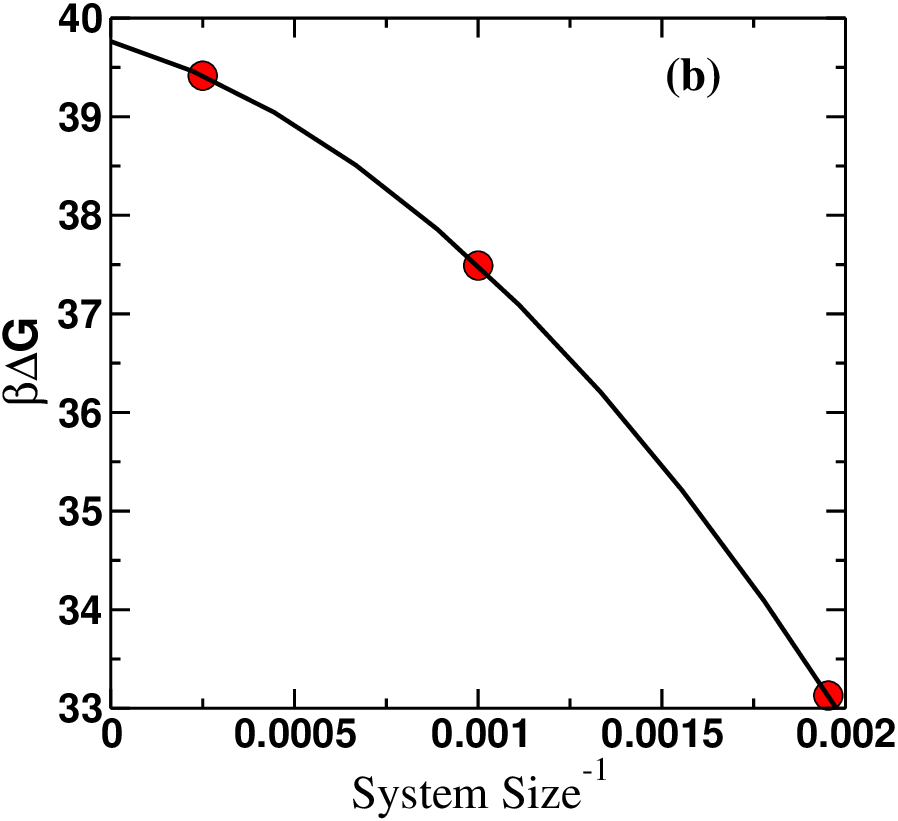}
\caption{Free energy calculations performed using USMC runs with a harmonic bias on $n_{max}$ at $T=1221K,~P=0~GPa$ at $3$ values of the system size $N=512,1000,4000$. Additional runs with a hard wall bias at small $n_{max}$ are used to obtain improved sampling for small cluster sizes. We extrapolate the dependence of the barrier height on the system size and find that for $N=4000$, the free energy barrier approaches close to the asymptotic value for infinite system size.}
\label{fig:fig-d1}
\end{figure}

In addition to system size effects noted while using $n_{max}$ as the order parameter, related to the extensivity of $P(n_{max})$, another system size effect comes into play at state points where the critical cluster size is large. This is best understood by considering the fact that the density of a crystal is approximately $\rho_c=0.45\sigma^{-3}$ ($~\sim~2.3~gcc^{-1}$). We also know that $\rho_c=N/(l^3)=$ from which we get the box length for a given system size as $l=(\sigma^{3}N/0.45)^{1/3}$. For $N=512,1000,4000$ respectively this comes to approximately $10\sigma,13\sigma,21\sigma$. We now consider the radial extent of a crystalline cluster (assumed to be spherical) of size $n_{max}=80$. This is given by 
\begin{equation}
  r^3=\left(\frac{n_{max}}{\frac{4}{3}\pi\rho}\right)^{1/3}\sigma  
\end{equation}
For $n_{max}=80$ we get $r\approx3.5\sigma$. The diameter of this cluster will be greater than $l/2$ for $N=512$ and we can therefore expect the crystalline cluster to induce effects across periodic images. In order to avoid these effects, free energy calculations are best performed at $N=4000$ at state points where the critical cluster is large. This is illustrated in Fig.~\ref{fig:fig-d1} where we calculate the free energy at $T=1221K,~P=0~GPa$ at three system size.
\FloatBarrier
\bibliographystyle{aip}
\bibliography{nucleoref}
\end{document}